\PassOptionsToPackage{unicode}{hyperref}
\PassOptionsToPackage{hyphens}{url}
\documentclass[]{article}
\usepackage[margin=1in]{geometry}
\usepackage{xcolor}
\usepackage{amsmath,amssymb}
\usepackage{iftex}
\ifPDFTeX
  \usepackage[T1]{fontenc}
  \usepackage[utf8]{inputenc}
  \usepackage{textcomp}
\else
  \usepackage{unicode-math}
  \defaultfontfeatures{Scale=MatchLowercase}
  \defaultfontfeatures[\rmfamily]{Ligatures=TeX,Scale=1}
\fi
\usepackage{lmodern}
\IfFileExists{upquote.sty}{\usepackage{upquote}}{}
\IfFileExists{microtype.sty}{%
  \usepackage[]{microtype}
  \UseMicrotypeSet[protrusion]{basicmath}
}{}
\makeatletter
\@ifundefined{KOMAClassName}{%
  \IfFileExists{parskip.sty}{%
    \usepackage{parskip}
  }{%
    \setlength{\parindent}{0pt}
    \setlength{\parskip}{6pt plus 2pt minus 1pt}}
}{%
  \KOMAoptions{parskip=half}}
\makeatother
\usepackage{longtable,booktabs,array}

\usepackage{calc}
\usepackage{etoolbox}
\makeatletter
\patchcmd\longtable{\par}{\if@noskipsec\mbox{}\fi\par}{}{}
\makeatother
\IfFileExists{footnotehyper.sty}{\usepackage{footnotehyper}}{\usepackage{footnote}}
\makesavenoteenv{longtable}
\usepackage{graphicx}
\usepackage{framed}
\usepackage{fvextra}
\makeatletter
\newsavebox\pandoc@box
\newcommand*\pandocbounded[1]{%
  \sbox\pandoc@box{#1}%
  \Gscale@div\@tempa{\textheight}{\dimexpr\ht\pandoc@box+\dp\pandoc@box\relax}%
  \Gscale@div\@tempb{\linewidth}{\wd\pandoc@box}%
  \ifdim\@tempb\p@<\@tempa\p@\let\@tempa\@tempb\fi%
  \ifdim\@tempa\p@<\p@\scalebox{\@tempa}{\usebox\pandoc@box}%
  \else\usebox{\pandoc@box}%
  \fi%
}
\def\fps@figure{htbp}
\makeatother
\usepackage[htt]{hyphenat}
\renewcommand{\texttt}[1]{{\ttfamily\hyphenchar\font=`\-\relax #1}}
\providecommand{\tightlist}{%
  \setlength{\itemsep}{0pt}\setlength{\parskip}{0pt}}
\definecolor{shadecolor}{RGB}{248,248,248}
\newenvironment{Shaded}{\begin{snugshade}}{\end{snugshade}}
\DefineVerbatimEnvironment{Highlighting}{Verbatim}{commandchars=\\\{\},breaklines,breakanywhere}
\newcommand{\NormalTok}[1]{#1}
\usepackage{bookmark}
\IfFileExists{xurl.sty}{\usepackage{xurl}}{}
\hypersetup{
  hidelinks,
  pdfcreator={LaTeX via pandoc}}

\title{Digital Identity for Agentic Systems: Toward a Portable\\Authorization Standard for Autonomous Agents}
\author{Partha Madhira}
\date{}

\newif\ifwatermark
\IfFileExists{watermark.flag}{\watermarktrue}{\watermarkfalse}
\ifwatermark
  \usepackage{draftwatermark}
  \SetWatermarkText{P. Madhira - DRAFT}
  \SetWatermarkScale{0.5}
  \SetWatermarkLightness{0.85}
\fi

\begin{document}
\maketitle
\section*{Abstract}

Enterprise AI is shifting from copilots to autonomous agents capable of
executing workflows, negotiating outcomes, and making decisions with
limited human oversight. While many deployments remain internal today,
increasing pressure exists to extend these systems across organizational
boundaries, for example, through emerging initiatives such as MIT's
Project NANDA, which explore the foundations of an ``Internet of
Agents'' {[}20{]}. This shift introduces new trust challenges: not only
must an agent be identifiable, but its authority must also be explicit,
constrained, auditable, and revocable.

This paper analyzes two representative enterprise use cases: insurance
claims processing and supply chain integrity in defense and aerospace.
It uses them to surface the identity and trust requirements that emerge
when autonomous agents operate across organizational boundaries. It
first examines structural gaps in existing identity models through these
scenarios, and then proposes a portable authorization standard for
autonomous agents: a common authorization semantic model with typed
constraint algebra, decision-consistent evaluation semantics, delegation
attenuation rules, governed semantic resolution, and pre-flight
discovery. It also outlines a layered standardization path that
separates a normative core from profile bindings, governed vocabularies,
and optional advanced governance profiles for stateful enforcement.

\subsection{1. Introduction}\label{introduction}

Artificial intelligence systems are undergoing a structural transition
from tools that assist human users to autonomous agents capable of
initiating and executing impactful actions. In enterprise environments,
this evolution is particularly significant. Agents are no longer
confined to generating recommendations or summarizing information; they
are increasingly entrusted with tasks such as approving transactions,
coordinating critical workflows, and interacting with internal and
external systems on behalf of organizations.

This shift introduces a trust problem regardless of whether agents are
coordinating workflows internally within an enterprise or submitting
actions to third-party vendors, suppliers, service providers, or
clients. Even within a single organization, identity, access control,
and audit responsibilities are often distributed across business units,
control functions, and operational systems. As a result, accountability
is not always straightforward to establish when autonomous agents begin
to act across these internal lines of control.

An example workflow within an enterprise is an IT operations agent
autonomously placing an order with an external vendor without involving
procurement. While the action may technically originate from within the
enterprise, it bypasses established financial and procedural control
points. The immediate concern is not merely whether the agent was
authenticated, but whether the agent had bypassed the established
process, potentially exposing the enterprise to audit failure,
non-compliance, or financial loss. This requires answering questions
through conscious, well-considered design about authorization, policy
compliance, and auditability. All of these concerns already exist within
an enterprise even before integrations with external parties come into
play.

These gaps may sometimes remain contained, tolerated, or compensated for
within enterprise boundaries. They become significantly more pronounced
and consequential, however, when agent interactions extend across
organizational boundaries to partners, suppliers, service providers,
customers, and regulators. In such settings, each party operates under
its own policies, contractual obligations, and accountability regimes.
Actions may carry legal, financial, or regulatory consequences, and
failures can propagate across ecosystems rather than remaining confined
to a single control environment.

Seen in this light, the central challenge is not defined by where an
agent acts, but by how its authority is established, interpreted, and
governed as it acts. Similar questions arise regardless of whether the
action is internal or cross-boundary: what authority does the agent
hold, who granted it, under what conditions may it act, and how can its
actions be attributed and audited over time?

Existing identity and access management models were developed primarily
to authenticate principals and authorize access to resources, whether
for human users or machine workloads. They remain effective for many
enterprise and cross-organizational integrations. However, they provide
limited standardized support for expressing portable, issuer-authored
authorization constraints that can be evaluated consistently as
autonomous actions unfold over time under changing conditions.

The discussion that follows examines how digital identity must evolve as
AI systems shift from read-only analysis to autonomous action-taking. It
analyzes representative enterprise scenarios to surface the identity and
trust requirements that emerge when autonomous agents operate across
trust boundaries. In doing so, it explores how identity must move beyond
point-in-time authentication toward supporting the evaluation and
enforcement of delegated authority as agent actions unfold.

\subsection{2. Background and Structural
Limitations}\label{background-and-structural-limitations}

Traditional identity and access management systems are built on a
relatively simple model: a subject is authenticated, and based on that
authentication, access to resources is granted or denied. Mechanisms
such as OAuth tokens, API keys, and role-based access control have
proven effective in environments where interactions are short-lived,
well-defined, and largely synchronous.

However, these models assume that access decisions can be made at the
moment a request is initiated. They do not account for workflows that
unfold over time, nor do they capture the intent or context behind a
sequence of actions. In agentic systems, this limitation becomes
critical. An agent may initiate a process, make intermediate decisions,
interact with multiple parties, and eventually drive an outcome, all
without direct human intervention at each step.

Simple bearer-token-based authorization exemplifies this gap. A bearer
token can confirm that a system is permitted to access a resource, but
it provides limited information about whether a specific action is
appropriate within a given business context. More expressive
authorization frameworks can carry richer claims, delegated context, or
structured authorization details, but they still do not generally define
a portable evaluation model for issuer-authored constraints such as
bounded settlement authority, restrictions limiting an agent to approved
recipients or data domains, validity only within a defined temporal
window, or cumulative exposure limits, for example, authority to place
individual orders up to USD 2000, but no more than USD 50,000 in
aggregate within a week. As a result, systems often compensate through
centralized orchestration, bilateral integration logic, or post hoc
auditing.

These workarounds become increasingly fragile as autonomy and
distribution increase. When agents operate across organizational
boundaries, each interaction carries potential legal, financial, or
regulatory implications. The distinction between authentication and
authorization becomes more pronounced: it is no longer sufficient to
know who is acting; it is necessary to determine whether the action
itself is justified, compliant, and within the bounds of delegated
authority.

Existing models therefore exhibit a structural limitation. They are
designed to answer whether access should be granted at a point in time,
but they are not designed to govern how authority is exercised across a
sequence of decisions. In agentic systems, however, risk does not arise
solely at the initiation of an interaction, but throughout the chain of
actions that follow.

This mismatch introduces a fundamental challenge. Identity must evolve
from a mechanism that gates access to one that supports the evaluation
and enforcement of authority as actions unfold over time. It must be
capable of expressing not only who an agent is, but also what it is
allowed to do, under what constraints, and with what accountability.

\subsection{3. Analytical Framework}\label{analytical-framework}

The analysis centers on concrete enterprise scenarios in which
autonomous agents operate across organizational boundaries within and
across enterprises. These scenarios provide a practical basis for
identifying where existing identity models remain sufficient, where they
begin to fail, and what additional requirements emerge when agents are
allowed to act rather than merely assist.

Each use case is examined along three dimensions. First, the relevant
actors and trust boundaries are identified, with particular attention to
where authority, data, and responsibility cross organizational lines.
Second, the interactions between agents are examined in detail, focusing
on the sequence of actions, decision points, and dependencies that
characterize the workflow. Third, the identity and authorization
requirements that emerge from these interactions are articulated, with
an emphasis on what must be verifiable, enforceable, and auditable.

These three dimensions are used because they expose different layers of
the same problem. Actors and trust boundaries reveal where
accountability becomes fragmented. Agent interactions reveal how risk
materializes through the unfolding chain of actions rather than at a
single moment of access. Identity and authorization requirements reveal
what existing systems must be able to represent, evaluate, and enforce
if those interactions are to remain trustworthy. Taken together, these
dimensions make it possible to move from isolated examples to a more
general understanding of the structural gaps introduced by action-taking
agents.

The use cases presented here insurance claims processing and supply
chain integrity in defense and aerospace, are not drawn from any single
source. They are enterprise-oriented analytical scenarios informed by
years of experience in production-grade implementation. They are not
intended as an exhaustive catalogue of agentic systems. Several
candidate scenarios were considered, and these were selected because,
taken together, they surface the principal failure modes discussed in
this paper across both operational decision-making and multi-party
coordination settings. They are used as representative settings in which
authority, verification, policy compliance, and auditability are all
stressed in ways that make underlying identity limitations easier to
observe.

\subsection{4. Use Case 1: Agentic Insurance Claims
Processing}\label{use-case-1-agentic-insurance-claims-processing}

Insurance claims processing provides a representative example of how
autonomous agents may operate across organizational boundaries. The
process involves multiple independent parties, including policyholders,
insurers, adjusters, service providers, and reinsurers, each operating
under distinct contractual and regulatory obligations.

In an agentic model, each of these parties may be represented by
software agents capable of performing tasks such as collecting evidence,
validating coverage, coordinating service providers, and executing
payments. A typical workflow begins when a policyholder, or an agent
acting on their behalf, submits a claim. The insurer's agent verifies
policy details, evaluates submitted evidence, and determines the
appropriate processing path.

As the workflow progresses, interactions extend beyond the insurer's
internal systems. An adjuster agent may request repair estimates from
external service-provider agents, which must in turn demonstrate their
legitimacy and compliance with required standards. A negotiator agent
may also interact with the policyholder to negotiate the adjustment
amount, as commonly occurs in claims handling. Such an agent would
require clearly bounded authorization, including guidance on the range
within which it may negotiate and when it must escalate to the adjuster
or another authority. In more complex cases, reinsurer agents may be
involved to assess exposure and confirm coverage thresholds. Finally, a
payment agent may execute disbursement once all conditions are
satisfied.

These interactions are not merely data exchanges; they are
decision-bearing steps that carry financial and legal implications. Each
participating agent must therefore be able to establish not only its
identity, but also the authority under which it operates.

From this scenario, several identity and trust requirements emerge.
Agents must possess identities that are verifiable across organizational
boundaries, enabling counterparties to establish trust without relying
solely on bilateral integrations. Authority must be explicitly
delegated, with clear constraints on scope, duration, and permissible
actions. Policies governing data sharing, regulatory compliance, and
decision thresholds must be enforceable as part of the interaction,
rather than being applied only at system boundaries.

Additionally, the entire workflow must be auditable. Autonomous
decisions must be traceable to specific agents, the authority under
which they acted, and the evidence they considered. Finally, delegation
must be revocable, allowing organizations to withdraw authority in
response to errors, misconfigurations, or security incidents without
disrupting the broader system. This may become necessary, for example,
when an agent repeatedly attempts to query a database beyond its
expected scope or tries to copy more information than is required for
the task at hand, indicating either malfunction, overreach, or
compromise.

Current approaches struggle to meet these requirements. Centralized
workflow engines and bearer-token authorization models are effective for
tightly coupled systems but provide limited support for expressing
delegated authority or enforcing constraints across distributed,
asynchronous interactions. As a result, they rely heavily on predefined
integrations and retrospective auditing, which only addresses the
aftermath of an action rather than giving humans a meaningful
opportunity to prevent it from occurring in the first place, and
therefore becomes insufficient as autonomy increases.

\begin{center}
\includegraphics[width=0.9\textwidth]{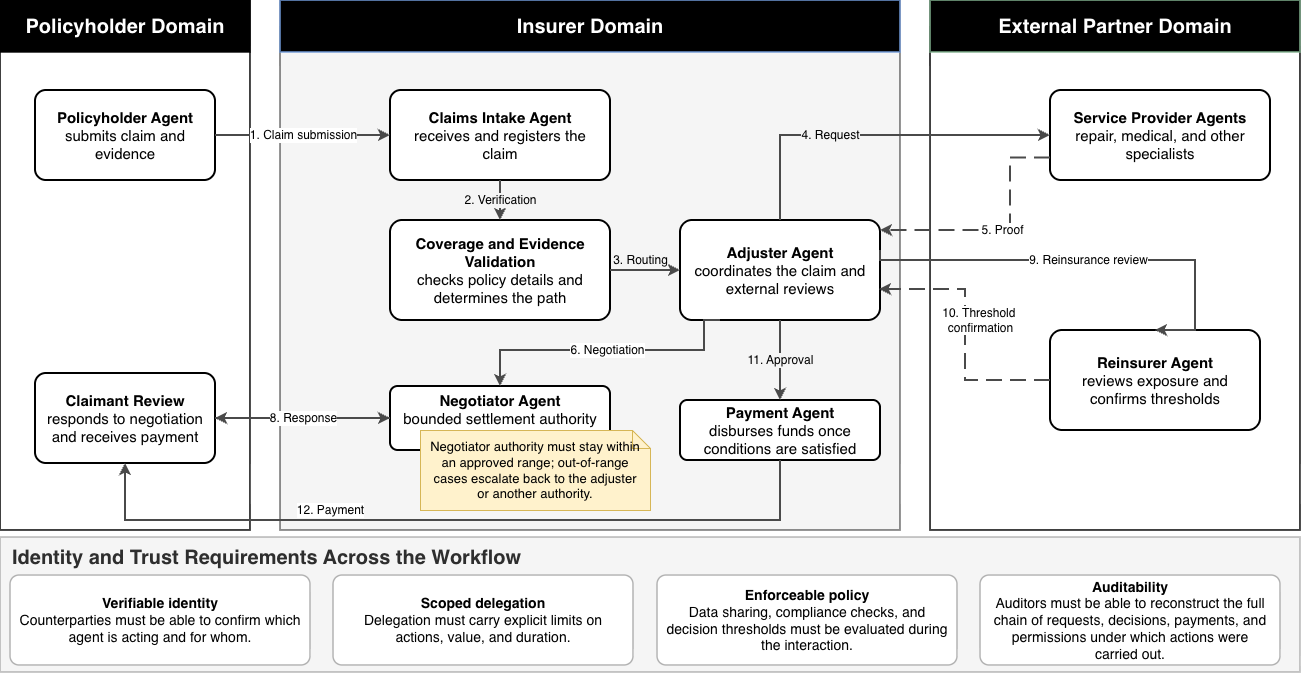}
\end{center}

\emph{Figure 1: Agent Insurance Claims Processing use case:
representative cross-organizational workflow showing how agents verify
coverage, coordinate external parties, negotiate settlements, involve
reinsurers when needed, and execute payment.}

\subsection{5. Use Case 2: Agentic Supply Chain Integrity in Defense and
Aerospace}\label{use-case-2-agentic-supply-chain-integrity-in-defense-and-aerospace}

A second representative scenario arises in high-assurance supply chains
such as those found in defense and aerospace manufacturing. In these
environments, components pass through multiple tiers of suppliers before
they are assembled into systems whose failure may carry severe
operational, financial, or safety consequences. A defect in a single
part may trigger costly investigations, production delays, certification
issues, or downstream failures.

In such settings, enterprises are increasingly interested in digital
continuity across the lifecycle of a part. The component or product
identifier functions here as the stable business reference for a
long-running cross-boundary workflow, while the agent carries the
authorization needed to request, relay, or present evidence associated
with that reference. An autonomous agent may follow a component as it
moves from one supplier to another, collecting attestations about its
provenance, material properties, testing history, and compliance with
required specifications. As the part moves from a lower-tier
manufacturer to an assembler, the agent must preserve a chain of
evidence that can later be inspected by auditors, regulators, or prime
contractors.

These interactions introduce a distinct trust challenge. Each supplier
may need to prove that its contribution satisfies contractual or
technical requirements without disclosing proprietary manufacturing
processes, sensitive operational details, or the identities of its own
sub-suppliers. At the same time, downstream parties must be able to
determine whether the presenting agent is authorized to carry, relay, or
request specific attestations, and whether those attestations can be
relied upon in later certification or audit processes.

The identity requirements that emerge from this use case are
substantial. Agents must be identifiable across organizational
boundaries and must be able to demonstrate the authority under which
they are requesting, carrying, or presenting lifecycle evidence.
Authorization must be specific enough to govern what types of
attestations an agent may access or transmit, under what conditions, and
to which parties. Auditability is also essential, since failures often
require reconstruction of the chain of responsibility across many
organizations.

Current approaches rely heavily on manual attestations, fragmented
documentation, and delayed reconciliation of records. These mechanisms
are slow, brittle, and poorly suited to machine-speed coordination
across large supplier networks. They also make it difficult to
determine, in real time, whether an agent is acting within its permitted
scope or whether a presented chain of evidence remains trustworthy as it
moves across organizational boundaries.

\begin{center}
\includegraphics[width=0.9\textwidth]{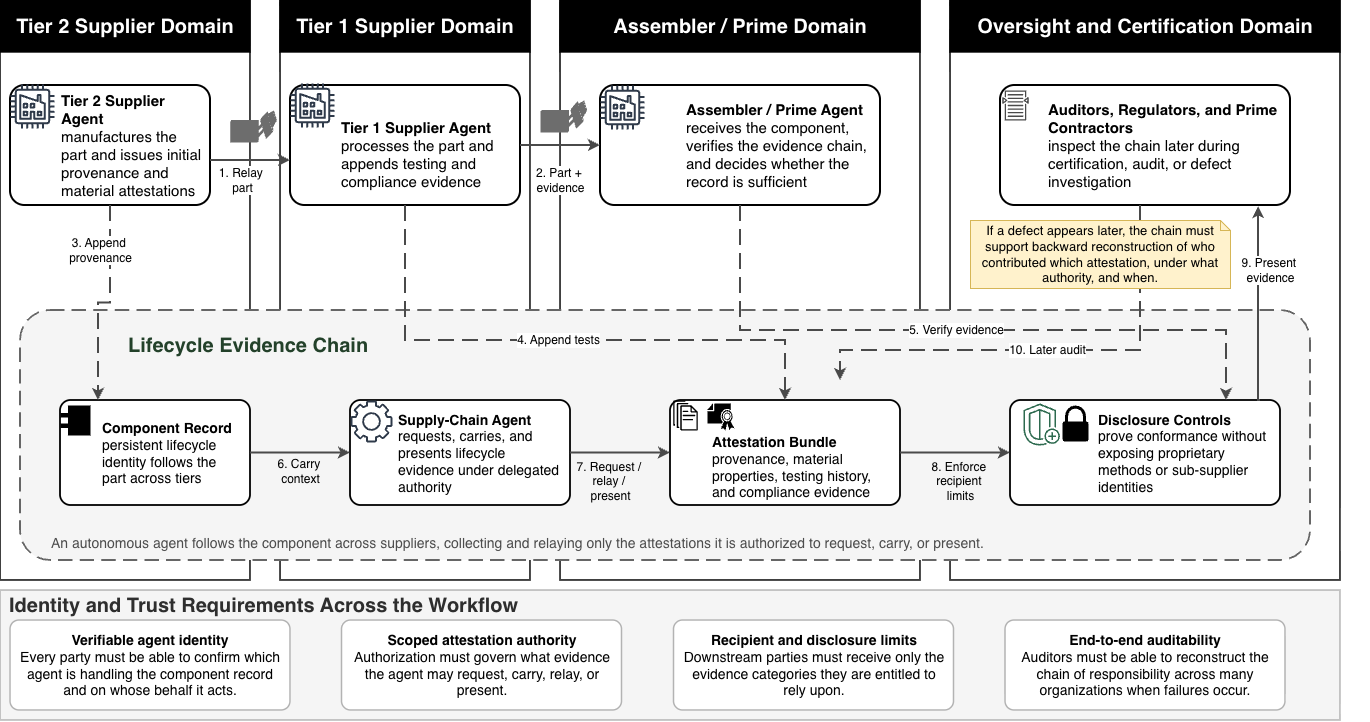}
\end{center}

\emph{Figure 2: Agentic Supply Chain Integrity in Defense and Aerospace
use case: a component moves across supplier tiers while an autonomous
agent preserves provenance, test, and compliance evidence for downstream
certification, audit, and investigation.}

\subsection{6. Cross-Case Synthesis}\label{cross-case-synthesis}

The two use cases examined in this paper: agentic insurance claims
processing and agentic supply chain integrity in defense and aerospace
arise in different industries, operate under different accountability
regimes, and involve different constellations of actors. Yet they
surface a remarkably consistent set of identity and trust requirements.
This section synthesizes those requirements and identifies the
architectural gap that neither existing identity standards nor emerging
agent communication protocols currently address.

\subsubsection{6.1 Five Recurring
Requirements}\label{five-recurring-requirements}

Five identity-related capabilities recur across both use cases.

\begin{enumerate}
\def\labelenumi{\arabic{enumi}.}
\item
  \textbf{Delegated authority for non-human actors.} In both domains,
  software agents must act with explicit, bounded authority on behalf of
  organizations or individuals. An insurer's negotiator agent must carry
  proof that it is authorized to settle within a defined range and under
  clearly bounded conditions. A supply-chain agent must carry proof that
  it is authorized to request, relay, or present specific attestations
  as a component moves across supplier tiers. In each case, authority
  must be scoped by purpose, bounded by constraints, and revocable
  without disrupting the broader system.
\item
  \textbf{Verifiable agent identity across organizational boundaries.}
  Internal agent deployments can often rely on platform-level mechanisms
  such as service accounts or workload identity. Across organizational
  boundaries, however, the requirement is stronger: counterparties must
  be able to verify not only that the presenting agent is authentic, but
  also that it is acting under legitimate, bounded authority. Federated
  workload identity can help establish strong presenter authentication
  across trust domains, but it does not by itself convey portable,
  delegated business authorization with bounded constraints, audit
  semantics, and revocation behavior that independent parties can
  interpret consistently. Thus, when an insurer's adjuster agent
  requests evidence from an external service-provider agent, or when a
  prime contractor receives lifecycle evidence from a lower-tier
  supplier's agent, each party must be able to determine that the other
  is both genuine and appropriately authorized, rather than an
  impersonator or a compromised system.
\item
  \textbf{Policy-bound authorization rather than broad, bearer-based
  access.} Bearer tokens are the dominant authorization artifact in
  modern API architectures. They grant access but carry little
  information about what the holder is specifically permitted to do,
  under what constraints, or for how long. Both use cases require
  authorization that reflects business and regulatory rules: monetary
  ceilings, negotiation ranges, temporal windows, resource restrictions,
  disclosure boundaries, recipient limitations, and escalation
  thresholds. These rules must travel with the agent, not reside solely
  in the receiver's configuration.
\item
  \textbf{End-to-end auditability of autonomous decisions.} When an
  agent negotiates a settlement, requests or relays a material
  certificate, appends an attestation to a chain of evidence, or
  approves a downstream action, the interaction must be traceable.
  Auditors, regulators, certification bodies, and dispute-resolution
  processes require the ability to reconstruct what happened, who
  authorized it, under what constraints, and why the system produced the
  outcome it did. This requirement is not merely operational; in
  regulated and high-assurance industries, it is often a legal or
  certification obligation.
\item
  \textbf{Revocation and accountability.} If an agent is compromised,
  misconfigured, or exceeds its mandate, its authority must be withdrawn
  promptly. When errors occur, responsibility must be attributable to
  the organization that empowered the agent. Revocation must propagate
  through delegation chains: revoking a parent credential must
  invalidate downstream delegations derived from it.
\end{enumerate}

\subsubsection{6.2 Different Facets of the Same
Problem}\label{different-facets-of-the-same-problem}

While both use cases converge on the same five requirements, they stress
different facets of the underlying model.

Insurance claims processing exercises \textbf{runtime constraint
evaluation and delegation attenuation} most heavily. The workflow
involves a chain of specialist agents such as claims intake, fraud
analysis, valuation, negotiation, and payment, each operating under
progressively narrower authority. A claims authority delegates to an
orchestrator, which delegates to a specialist, which may delegate to a
sub-specialist. At each hop, the scope must narrow or hold constant; it
must never widen. The receiving party evaluates the embedded constraints
against the live request context in near-real time: is the claim amount
within the authorized ceiling, is the negotiation range still in bounds,
is the claim type permitted, is the request within the allowed
processing window? The tempo is fast: minutes to hours, and the
evaluation is point-in-time.

The defense and aerospace supply-chain use case exercises
\textbf{chain-of-evidence integrity and scoped attestation flow}. The
critical challenge is not only whether a single request falls within an
operational ceiling, but whether the agent's authority to request,
carry, or present specific categories of evidence can be verified,
whether the provenance of contributed attestations remains intact, and
whether the entire interaction can be reconstructed for regulatory,
contractual, or forensic review. The tempo is slower: hours to days or
longer, and the evidentiary chain is longer and must withstand scrutiny
months or years later.

This contrast is instructive. A model that handles only runtime
constraint evaluation would serve insurance but leave supply-chain
integrity underserved. A model that handles only chain-of-evidence
integrity would serve supply chains but miss the real-time enforcement
requirements of claims processing. The architectural challenge is to
define a common authorization semantic model that accommodates both:
structured, machine-evaluable constraints for real-time enforcement,
combined with cryptographic provenance and delegation chain integrity
for post-hoc audit.

\subsubsection{6.3 The Gap}\label{the-gap}

Existing standards and emerging protocols address fragments of this
problem, but the challenge is one of composition and interoperation
across trust boundaries, not simply the absence of individual building
blocks.

\textbf{W3C Verifiable Credentials} {[}7{]} define a data model for
tamper-evident, cryptographically verifiable claims. They establish who
issued a credential, who it is about, and when it expires. They do not
define how to embed machine-evaluable authorization constraints within a
credential, how those constraints should be evaluated at runtime, or how
delegation should attenuate authority across a chain.

\textbf{OAuth 2.0 and its extensions} {[}2{]}-{[}5{]} provide a mature
framework for delegated access to resources across both user-mediated
and machine-to-machine interactions. In practice, however, OAuth leaves
substantial discretion to receiving systems in how authorization
semantics are modeled and enforced. Conventional scopes are often
coarse-grained, and even more expressive mechanisms such as structured
authorization details improve request specificity more than
cross-receiver semantic portability. As a result, two conformant
receivers may still interpret the same authorization artifact
differently at the level of business policy. OAuth is therefore highly
effective for resource access delegation, but it does not by itself
define a portable evaluation model that independent receivers must apply
deterministically to issuer-authored constraints.

\textbf{MCP-related identity and agent credential efforts} {[}17{]},
including those arising around agent communication frameworks, improve
the representation of agent identity, roles, and delegated assertions.
However, they do not yet establish a common evaluation model for
issuer-authored authorization constraints, merge behavior, and
attenuation semantics that independent receivers can be expected to
interpret consistently. As a result, interoperability at the level of
authentication or credential exchange does not by itself guarantee
interoperable authorization outcomes.

\textbf{Capability-based models} such as UCAN {[}11{]}, Biscuit
{[}12{]}, and zCAP-LD {[}23{]}, {[}24{]} provide strong support for
delegation, attenuation, and cryptographic chaining. zCAP-LD is
especially notable for making the separation between a capability
envelope and pluggable caveats explicit. However, zCAP-LD remains an
incubating W3C CCG work item and is therefore still subject to change.
If and when it matures into a stable standard, this model aspires to
leverage its delegation and attenuation mechanics cleanly. The center of
gravity of these standards, however, is capability transfer rather than
a standardized constraint algebra for enterprise-style authorization
conditions such as monetary bounds, temporal windows, recipient
restrictions, or cumulative exposure controls. They therefore address an
important part of the problem, but not the full portable evaluation
model proposed here.

\textbf{Policy engines} such as OPA/Rego {[}14{]}, Cedar {[}15{]}, and
Zanzibar {[}16{]} demonstrate that fine-grained authorization can be
enforced effectively at runtime. Their primary design center, however,
is receiver-defined policy evaluation. They do not by themselves
standardize a portable, issuer-authored authorization payload with
shared semantics across independent receivers, which is the gap
addressed here.

\textbf{OpenID AuthZEN} {[}21{]} is directly relevant because it
standardizes an authorization API between Policy Enforcement Points and
Policy Decision Points. It improves interoperability for how
authorization decisions are requested and returned, including access
evaluation, batched evaluation, search, and PDP metadata discovery. The
model proposed here addresses a different layer: the portable,
issuer-authored authorization payload that an agent presents across
trust boundaries, together with typed constraints, semantic resolution,
delegation attenuation, and deterministic receiver-side evaluation
semantics. In practice, the two approaches are complementary: a receiver
could expose or consume AuthZEN-style PDP interfaces while evaluating
the portable authorization credential defined in this paper.

The gap, stated precisely, is the absence of a widely adopted,
container-agnostic authorization model that combines issuer-authored,
machine-evaluable constraints with defined evaluation semantics,
delegation attenuation, most-restrictive-wins composition, and
cryptographically anchored auditability, while enabling independent
receivers to reach consistent decisions from the same signed
authorization and request context.

\begin{center}
\includegraphics[width=0.9\textwidth]{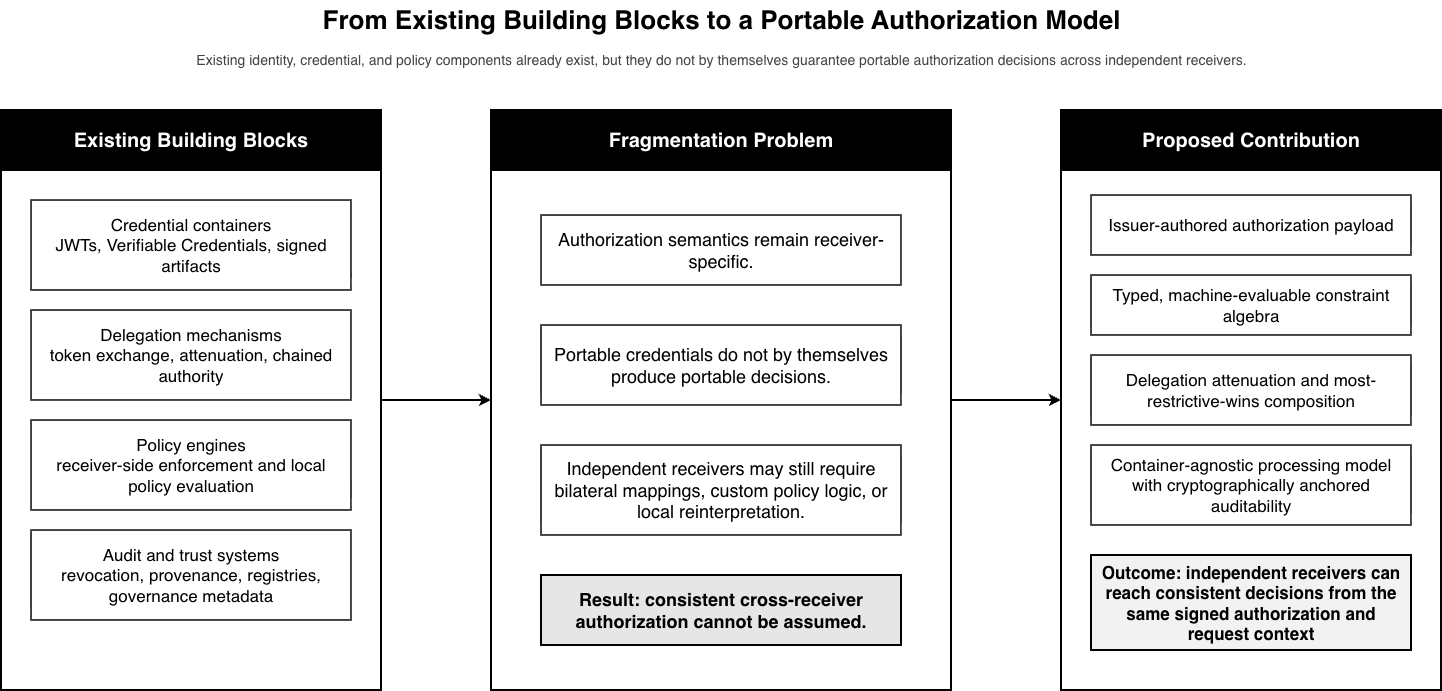}
\end{center}

\emph{Figure 3: The proposed contribution is not the credentials,
delegation, policy engines, or audit in isolation; it is the integration
of these elements into a portable authorization model with shared
evaluation semantics across independent receivers.}

\subsection{7. A Common Authorization Semantic Model for Autonomous
Agents}\label{a-common-authorization-semantic-model-for-autonomous-agents}

This section presents the core technical contribution of the paper: a
formal authorization semantic model for autonomous agents operating
across organizational boundaries. The model is intentionally separated
from any specific credential format or policy engine. It defines what
the authorization payload must contain, how constraints are structured
and evaluated, how delegation attenuates authority, and how the
processing pipeline produces a deterministic allow-or-deny decision for
any given input.

\subsubsection{7.1 Architectural Separation: Container, Payload,
Engine}\label{architectural-separation-container-payload-engine}

The first design principle is a strict three-layer separation.

\textbf{Layer 1 --- Credential Container.} The cryptographic envelope
that carries the authorization payload, binds it to an issuer identity,
and provides tamper evidence. This may be a W3C Verifiable Credential
{[}7{]}, a standard JWT {[}1{]} with structured claims, or another
signed container. The container provides integrity and authenticity. It
does not define what the authorization means.

\textbf{Layer 2 --- Authorization Payload.} The structured,
machine-evaluable content that defines what an agent is authorized to do
and under what conditions. This is the normative core of the model. It
is encoding-independent: the same semantic content can be serialized
into a VC field, a JWT claim set, or another supported representation.
For consistency, this paper refers to the signed, portable unit that
carries this payload as an \textbf{authorization credential}. The term
\textbf{artifact} is reserved for related outputs such as audit records,
discovery manifests, and state vouchers. What matters is not the
serialization but the presence of a defined set of mandatory components
and typed constraints.

\textbf{Layer 3 --- Enforcement Engine.} The runtime system that
verifies the credential container, extracts the authorization payload,
evaluates the constraints against the actual request context, merges
with local policy, and produces a decision. This may be OPA, Cedar,
Zanzibar, a purpose-built evaluator, or another runtime that implements
the processing pipeline defined below. The engine is interchangeable;
the evaluation semantics are not.

\textbf{Note on Terminology:} This model uses the terms Container,
Payload, and Engine to define a functional architecture for portable
authorization. While these concepts overlap with existing frameworks
(like XACML, OAuth, or ToIP), they are defined here as distinct,
independent logical layers to ensure that the authorization payload
remains the stable, deterministic contract regardless of the underlying
trust infrastructure or runtime stack.

This separation is analogous to the relationship between a declarative
specification and the runtime that enforces it. Two different
enforcement engines that both correctly implement the evaluation rules
defined here should produce the same authorization decision from the
same credential and request context. The practical consequence is that
the standard operates at the payload layer. It defines the grammar, that
is, the constraint types, the evaluation algebra, the attenuation rules,
and the processing pipeline. It does not mandate a specific container or
engine.

This three-layer model is a problem-specific decomposition of portable
agent authorization, not a restatement of a broader trust-stack
taxonomy. Governance artifacts such as manifests, trust registries, and
mapping profiles remain important, but in this paper they function as
external inputs and constraints on authorization processing rather than
as a separate runtime layer in the figure. Layer 3 is a logical
evaluation function, not a claim about deployment topology. A conformant
implementation may realize it through a unified evaluator, a PEP/PDP
split, or another internal architecture, provided the verification and
decision semantics defined here are preserved. Pure reference tokens,
API keys, and identity-only credentials that cannot support carrying
cryptographically bound authorization credentials naturally remain out
of scope.

For clarity, this paper uses the following terms consistently. The
\textbf{authorization payload} is the machine-evaluable semantic
content. The \textbf{authorization credential} is the signed portable
unit formed when that payload is bound into a credential container. The
\textbf{issuer} is the authority that grants that credential. The
\textbf{subject agent} is the agent to whom the credential applies. The
\textbf{presenting agent}, or \textbf{presenter}, is the runtime actor
that presents the credential; \textbf{subject binding} requires the
presenting agent to match the subject agent. The term \textbf{artifact}
is reserved for related outputs such as audit records, discovery
manifests, and state vouchers. Where the term \textbf{enforcer} appears
in advanced stateful profiles, it refers to a receiver-side enforcement
engine instance.

\begin{center}
\includegraphics[width=0.9\textwidth]{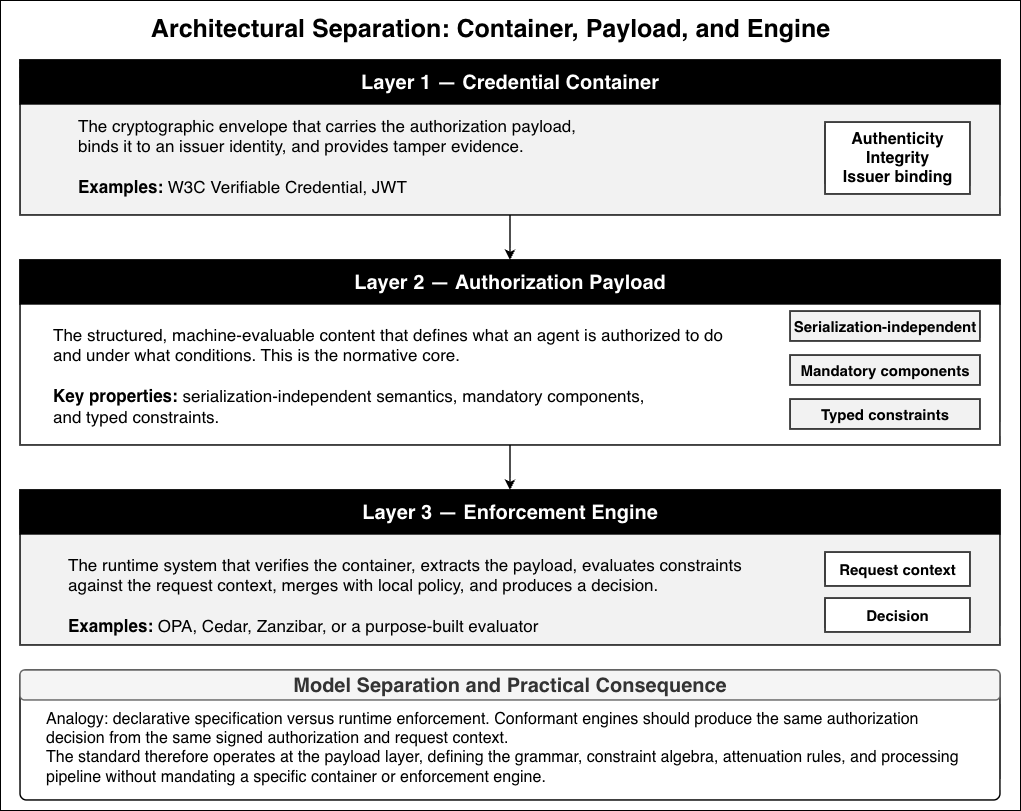}
\end{center}

\emph{Figure 4: Three-layer authorization architecture separating the
credential container, authorization payload, and enforcement engine.}

\subsubsection{7.2 Authorization Payload: Four Mandatory
Components}\label{authorization-payload-four-mandatory-components}

Every agent authorization credential, regardless of container format,
must carry four components.

\textbf{Component 1 --- Agent Identity.} The identifier of the agent to
whom this authorization applies. This binds the credential to a specific
agent; no other agent can legitimately present it.

\textbf{Component 2 --- Issuer Identity.} The identifier of the entity
that granted this authorization. This is the delegator, which may be a
human, an enterprise IAM system, or another authorized agent acting
within its own delegated authority. The issuer identity must be
cryptographically verifiable through the container's signature.

\textbf{Component 3 --- Declared Permissions.} What the agent is allowed
to do. These are expressed as explicit action identifiers. If the agent
requests an action not covered by its declared permissions, the request
is denied regardless of whether the constraints would otherwise pass.

\textbf{Component 4 --- Policy Constraints.} The machine-evaluable
conditions under which the declared permissions apply. These are the
enforceable rules embedded in the credential: monetary ceilings,
temporal windows, resource patterns, categorical restrictions, recipient
limitations, or other conditions that can be evaluated at runtime
against the actual request context.

Together, these four components constitute the portable authorization
policy. The policy is signed as a unit by the issuing authority. It
cannot be modified after issuance without invalidating the signature. It
can only be narrowed through delegation, never widened.

\subsubsection{7.3 Typed Constraint
Algebra}\label{typed-constraint-algebra}

The model defines four core constraint types. Each type has defined
properties, evaluation rules, and delegation attenuation rules. The
choice of four types, rather than an unconstrained expression language,
is deliberate and reflects a design tradeoff between expressiveness and
determinism. Taken together, these four types provide a minimum
deterministic basis for portable authorization across heterogeneous
receivers: scalar bounds, time validity, set membership, and namespace
scoping. More specialized behavior may be introduced through governed
profiles or advanced extensions, but the normative core remains
intentionally compact.

\paragraph{7.3.1 NumericLimitConstraint}\label{numericlimitconstraint}

A numeric limit constraint expresses bounded quantitative authority. It
contains a field name, an operator such as \texttt{eq}, \texttt{lt},
\texttt{lte}, \texttt{gt}, or \texttt{gte}, and a numeric value. It may
optionally include a unit or currency where relevant.

\textbf{Evaluation.} Extract the named field from the request context
and apply the operator against the declared value. If the comparison is
true, the constraint passes. If the field is absent or the value is not
numeric, the constraint fails. A bounded range may be expressed by
composing two numeric constraints over the same field; for example, a
floor expressed with \texttt{gte} and a ceiling expressed with
\texttt{lte} thereby supporting negotiation ranges without introducing a
separate range primitive into the core algebra.

\textbf{Example.} A negotiator agent may carry

{\small\noindent\texttt{NumericLimitConstraint(\\
\hspace*{2em}field=settlementAmount, operator=lte, value=5000, currency=USD)}}
.

\textbf{Attenuation rule.} A delegate's numeric limit must be equal or
more restrictive than the parent's. For ceiling operators such as
\texttt{lte}, the delegate's value must be less than or equal to the
parent's. A delegate cannot raise a ceiling.

\paragraph{7.3.2
TemporalWindowConstraint}\label{temporalwindowconstraint}

A temporal window constraint expresses when a permission is valid. It
includes a field, a valid-from time, a valid-until time, a timezone, and
optionally a set of allowed days.

\textbf{Evaluation.} Convert the request timestamp to the specified
timezone and confirm that it falls within the allowed interval. If the
field is absent, the constraint fails.

\textbf{Example.} A claims-handling credential may carry

{\small\noindent\texttt{TemporalWindowConstraint(\\
\hspace*{2em}field=requestTime, valid\_from=2026-04-01T00:00:00Z, valid\_until=2026-04-30T23:59:59Z, timezone=UTC)}}
.

\textbf{Attenuation rule.} A delegate's time window must be a subset of
the parent's. It may narrow the valid window but not broaden it.

\paragraph{7.3.3
EnumeratedListConstraint}\label{enumeratedlistconstraint}

An enumerated list constraint expresses bounded categories or recipient
sets. It references a field and defines either an allowed set, a denied
set, or both.

\textbf{Evaluation.} If an allowed set is present, the request value
must be a member. If a denied set is present, the value must not be a
member. If both apply, deny takes precedence. Missing fields fail
closed.

\textbf{Example.} A payment credential may carry
\texttt{EnumeratedListConstraint(field=recipientId,\ allowed={[}vendorA,\ vendorB,\ vendorC{]})}.

\textbf{Attenuation rule.} A delegate's allowed set must be a subset of
the parent's. A delegate's denied set may be equal or broader.

\paragraph{7.3.4 StringPatternConstraint}\label{stringpatternconstraint}

A string pattern constraint expresses bounded namespaces or resource
scopes. It references a field, a pattern, and a match type such as
\texttt{exact}, \texttt{prefix}, \texttt{suffix}, or a restricted glob.

\textbf{Evaluation.} Extract the field and apply the declared matching
rule. The \texttt{exact} match type requires the request value to match
the pattern literally. The \texttt{prefix} match type requires the
request value to begin with the declared string. The \texttt{suffix}
match type requires the request value to end with the declared string. A
restricted glob is an anchored wildcard pattern consisting only of
literal characters and the \texttt{*} wildcard, where \texttt{*} matches
zero or more characters. It excludes regular-expression features such as
character classes, alternation, grouping, backreferences, and
lookaround. Regular expressions are excluded from the core model to
avoid dialect fragmentation and security risks such as catastrophic
backtracking.

\textbf{Example.}

{\small\noindent\texttt{StringPatternConstraint(\\
\hspace*{2em}field=resourceId, match=exact, pattern=claims/CLM-2026-00412)}}

matches only that single resource.

{\small\noindent\texttt{StringPatternConstraint(\\
\hspace*{2em}field=resourceId, match=prefix, pattern=claims/)}}

matches any resource under the \texttt{claims/} namespace.

{\small\noindent\texttt{StringPatternConstraint(\\
\hspace*{2em}field=documentName, match=suffix, pattern=.pdf)}}

matches PDF artifacts.

{\small\noindent\texttt{StringPatternConstraint(\\
\hspace*{2em}field=resourceId, match=restricted\_glob, pattern=claims/*/attachments/*)}}

matches only attachment resources under the claims hierarchy.

\textbf{Attenuation rule.} A delegate's pattern must match a subset of
the strings matched by the parent's pattern. A delegate cannot broaden
the matched namespace.

\paragraph{7.3.5 Composite Constraint
Examples}\label{composite-constraint-examples}

The four core constraint types are intentionally simple, but they are
meant to be composed. In practice, an authorization credential will
often carry several constraints that are evaluated together under the
conjunctive fail-closed model described in Section 7.4.

\textbf{Example 1: insurance claims negotiator.} A claims negotiator
agent may be authorized to settle only certain claim types, within a
bounded negotiation range, during business-valid time windows, and only
for resources under the approved claims namespace.

\begin{Shaded}
\begin{Highlighting}[]
\NormalTok{permission: claim.settle}

\NormalTok{constraints:}
\NormalTok{  {-} NumericLimitConstraint(}
\NormalTok{        field=settlementAmount,}
\NormalTok{        operator=gte,}
\NormalTok{        value=2500,}
\NormalTok{        currency=USD}
\NormalTok{    )}
\NormalTok{    // Establishes the lower bound of the authorized negotiation range.}

\NormalTok{  {-} NumericLimitConstraint(}
\NormalTok{        field=settlementAmount,}
\NormalTok{        operator=lte,}
\NormalTok{        value=5000,}
\NormalTok{        currency=USD}
\NormalTok{    )}
\NormalTok{    // Establishes the upper bound; the agent cannot settle above USD 5,000.}

\NormalTok{  {-} TemporalWindowConstraint(}
\NormalTok{        field=requestTime,}
\NormalTok{        valid\_from=2026{-}04{-}01T00:00:00Z,}
\NormalTok{        valid\_until=2026{-}04{-}30T23:59:59Z,}
\NormalTok{        timezone=America/New\_York,}
\NormalTok{        allowed\_days=[Monday, Tuesday, Wednesday, Thursday, Friday]}
\NormalTok{    )}
\NormalTok{    // Limits settlement authority to the authorized calendar window and weekdays.}

\NormalTok{  {-} EnumeratedListConstraint(}
\NormalTok{        field=claimType,}
\NormalTok{        allowed=[auto\_collision, property\_damage]}
\NormalTok{    )}
\NormalTok{    // Restricts the agent to claim categories covered by the delegation.}

\NormalTok{  {-} StringPatternConstraint(}
\NormalTok{        field=claimId,}
\NormalTok{        match=prefix,}
\NormalTok{        pattern=claims/auto/}
\NormalTok{    )}
\NormalTok{    // Restricts the resource namespace to auto{-}claim records.}
\end{Highlighting}
\end{Shaded}

\textbf{Example 2: supply-chain evidence agent.} A supply-chain agent
may be authorized to present only approved evidence types, to approved
recipient roles, for components in a specific namespace, and with a
bounded disclosure volume.

\begin{Shaded}
\begin{Highlighting}[]
\NormalTok{permission: evidence.present}

\NormalTok{constraints:}
\NormalTok{  {-} EnumeratedListConstraint(}
\NormalTok{        field=evidenceType,}
\NormalTok{        allowed=[origin\_attestation, test\_certificate, material\_compliance]}
\NormalTok{    )}
\NormalTok{    // Allows only evidence categories that are approved for disclosure.}

\NormalTok{  {-} EnumeratedListConstraint(}
\NormalTok{        field=recipientRole,}
\NormalTok{        allowed=[prime\_contractor, certified\_auditor, regulator]}
\NormalTok{    )}
\NormalTok{    // Limits disclosure to recipient roles that are permitted to receive evidence.}

\NormalTok{  {-} StringPatternConstraint(}
\NormalTok{        field=componentId,}
\NormalTok{        match=restricted\_glob,}
\NormalTok{        pattern=aerospace/components/*/lot/*}
\NormalTok{    )}
\NormalTok{    // Restricts evidence presentation to components under a governed namespace.}

\NormalTok{  {-} TemporalWindowConstraint(}
\NormalTok{        field=requestTime,}
\NormalTok{        valid\_from=2026{-}01{-}01T00:00:00Z,}
\NormalTok{        valid\_until=2026{-}12{-}31T23:59:59Z,}
\NormalTok{        timezone=UTC}
\NormalTok{    )}
\NormalTok{    // Defines the period during which the evidence{-}presenting authority is valid.}

\NormalTok{  {-} NumericLimitConstraint(}
\NormalTok{        field=maxRecordsDisclosed,}
\NormalTok{        operator=lte,}
\NormalTok{        value=100}
\NormalTok{    )}
\NormalTok{    // Prevents excessive disclosure in a single interaction.}
\end{Highlighting}
\end{Shaded}

\paragraph{7.3.6 Extensibility}\label{extensibility}

The four core types are not a closed set. The model allows extension
types, but each extension must declare three things: its evaluation
semantics, its attenuation semantics, and its failure behavior. Any
constraint type not recognized by the evaluator must be treated as
failed. This preserves fail-closed behavior and prevents silent widening
of authority.

\subsubsection{7.4 Evaluation Semantics}\label{evaluation-semantics}

The evaluation model is conjunctive, total, and fail-closed.

\textbf{Conjunctive logic.} All constraints must pass for the request to
be allowed. Constraints are cumulative restrictions, not alternative
authorization paths.

\textbf{Fail-closed on unknown types.} If the evaluator encounters a
constraint type it does not recognize, the request is denied.

\textbf{Fail-closed on missing context.} If a constraint references a
field absent from the request context, the request is denied.

These semantics ensure that the evaluation is total: every well-formed
authorization credential against a request context produces a definite
allow-or-deny outcome. Two conformant implementations, given the same
signed authorization, request context, applicable profile and vocabulary
versions, and canonical typing and mapping rules, should therefore
produce the same authorization decision. It is this property of decision
consistency under shared semantics that makes the model standardizable.

\subsubsection{7.5 Most-Restrictive-Wins and Local
Policy}\label{most-restrictive-wins-and-local-policy}

The credential defines the authorization ceiling. The receiver's local
policy can lower that ceiling but never raise it. The effective
authorization is the intersection of the credential's constraints and
the receiver's local restrictions.

This is a familiar security principle. A network-level permit does not
override a more restrictive host-level policy. The same logic applies
here: an authorization credential authorizing claims up to a certain
threshold may still be narrowed further by the insurer's internal
control policy, or a supply-chain agent's relay permission may be
narrowed by a receiver's own disclosure rules.

The critical design constraint is that local policy should be
expressible using the same typed constraint model as the credential.
This keeps evaluation uniform and the resulting decision trace
auditable.

\paragraph{7.5.1 Multi-Principal Workflow
Composition}\label{multi-principal-workflow-composition}

Some enterprise workflows require more than one independently authorized
agent to contribute to a single outcome. A claims settlement, for
example, may require a claims adjuster agent to authorize the settlement
amount and a fraud detection agent to attest that fraud indicators are
below the receiver's threshold. These cases are not delegation chains:
neither agent necessarily derives authority from the other, and their
credentials may be issued by different authorities.

The model treats this as \textbf{multi-principal composition}, not
credential merging. Each credential is first evaluated independently
against the presenting agent, issuer trust, audience, semantic profile,
and request context. The workflow decision is then evaluated by the
receiver's workflow policy, which specifies which independently valid
credentials, roles, attestations, or approvals are required for the
outcome. A conformant evaluator must not combine credentials from
different issuers into a larger implied grant unless an applicable
profile explicitly defines such a composition rule.

Each authorization credential in the core model is itself conjunctive:
all constraints carried by that credential must pass for it to authorize
the requested action. The normative core does not define a separate
cross-credential algebra. When multiple credentials are presented, they
are evaluated independently; any requirement that several credentials
all be present and valid is defined by receiver workflow policy or by a
governed profile, not by nested Boolean operators in the credential
language itself.

The safe default is therefore conjunctive composition at the workflow
boundary: every required participant authorization must pass, and the
final workflow action must remain within the intersection of all
applicable constraints and local policy. If two credentials constrain
the same semantic field, the most restrictive compatible constraint
applies. If the constraints are incompatible, ambiguous, or scoped to
different meanings of the same field, the workflow must fail closed.
Union semantics, such as ``credential A allows amount up to USD 10,000
and credential B allows recipient X, therefore the combined system may
perform both,'' are not part of the core model unless explicitly
authorized by a governed workflow profile.

For example, a receiver may require one credential from a claims
adjuster agent authorizing \texttt{claim.settle} up to USD 5,000 and a
second credential from a fraud-screening agent attesting that the claim
is eligible for automated settlement. Each credential is evaluated
independently under the same fail-closed rules. The workflow policy then
requires both to be valid before the settlement proceeds. The core model
does not require a single composite credential to encode that
relationship, nor does it require a general cross-credential
\texttt{AND}, \texttt{OR}, or \texttt{XOR} algebra in the portable
payload.

\subsubsection{7.6 Three-Tier Trust
Gradient}\label{three-tier-trust-gradient}

The authorization semantic model is independent of how trust is
established between issuer and evaluator. In practice, however,
enterprises operate at different levels of trust infrastructure
maturity. The model accommodates three trust tiers.

\textbf{Tier 1 --- Bilateral.} Both parties already know each other. The
issuer publishes a JWKS endpoint {[}6{]} or exchanges signing keys
during partner onboarding. Authorization credentials may be standard
JWTs with embedded structured constraints.

\textbf{Tier 2 --- Federated.} The parties share a trust anchor but may
not have a direct bilateral relationship. X.509 certificates or
comparable PKI mechanisms provide the trust chain.

\textbf{Tier 3 --- Decentralized.} No prior relationship is required. A
verifiable credential carries its own trust chain, and the receiver
resolves the issuer identity through decentralized trust resolution.

The authorization semantics, the constraint types, evaluation algebra,
attenuation rules, merge behavior, and audit requirements are identical
across all three tiers. The trust tier is a deployment decision. The
authorization model is the standard.

The model therefore does not claim that one token class is universally
best for every deployment. A JWT/JWS profile is often the most practical
fit for current enterprise IAM and API ecosystems because it aligns with
existing tooling and lower operational overhead. A VC profile is useful
where portable issuer verifiability, decentralized trust resolution, or
richer credential lifecycles are required across trust boundaries, but
it can introduce additional privacy, status-management, and
implementation complexity. The purpose of the standard is to keep the
authorization meaning stable across those container choices rather than
to force ecosystem convergence on a single token type.

\paragraph{7.6.1 Trust Discovery and Governance
Frameworks}\label{trust-discovery-and-governance-frameworks}

In a decentralized environment, resolving an issuer's identity through a
DID or comparable decentralized mechanism is not sufficient to establish
business, regulatory, or domain-specific trust. Cryptographic
verification establishes authenticity: it can show that the credential
was issued by the controller of a given identifier and has not been
tampered with. It does not, by itself, establish that the issuer is
authorized to issue a particular class of authorization credential
within a given industry or governance regime. If the model is to support
zero-trust, cross-boundary autonomy at scale, it therefore requires a
standardized mechanism for trust discovery in addition to signature
verification.

\subparagraph{7.6.1.1 Trust Registry Model}\label{trust-registry-model}

In decentralized environments, the trust registry acts as a source of
truth for vetted trust anchors, situating the interaction within a
specific governance regime. Rather than serving as a static directory,
the registry provides the governance metadata required to validate the
agent's operating environment.

A conformant registry entry for agentic systems should provide at least
three categories of governance information.

First, it should establish \textbf{issuer standing}: whether the issuer
is an active and recognized authority for the specific domain in
question, such as insurance, aerospace, or another governed industry
profile.

Second, it should identify \textbf{permitted state authorities}: a
vetted set of state-store providers, registries, or protocols that the
issuer recognizes for cumulative governance. This prevents state
spoofing by ensuring that an enforcer does not update counters or
balances at an arbitrary location, but only at a state authority
pre-approved by the delegating governance framework.

Third, it should provide a \textbf{vocabulary reference}: governed
references to the vertical industry profiles or semantic vocabularies
the issuer uses for aliasing and profile interpretation, including
profiles aligned with initiatives such as Project NANDA {[}20{]} and A2A
Agent Cards {[}19{]} where applicable. These references are used for
onboarding, validation, and profile synchronization rather than as a
live runtime dependency for the evaluator.

These registries should be queryable through standardized trust
discovery frameworks or protocols, allowing evaluators to determine not
only whether an issuer is cryptographically real, but whether it is in
good standing for the relevant credential class, state authority model,
and domain vocabulary.

Nothing in this model requires authorization credentials or governance
metadata to be published to a public Verifiable Data Registry. Depending
on the deployment profile, the relevant trust registry or governance
source may be public, consortium-scoped, bilateral, or
enterprise-private, provided the evaluator can authenticate it and apply
its results deterministically.

\subparagraph{7.6.1.2 Integration with Phase 1
Verification}\label{integration-with-phase-1-verification}

This discovery process belongs in Phase 1 of container verification. The
\texttt{resolveIssuer} function should therefore be understood as
performing not only identifier resolution, but also governance
validation.

The extended verification sequence is as follows. First, the evaluator
extracts the issuer identifier from the presented artifact. Second, it
resolves the issuer's public key material through the applicable trust
mechanism. Third, it queries the relevant trust registry or governance
source to confirm that the issuer is an active and recognized authority
for the requested credential type, domain profile, or vocabulary
context. If the issuer cannot be found in a trusted registry, or if the
governance query fails in a way that leaves issuer standing unresolved,
the evaluator must fail closed and deny the request.

This distinction is important because it separates two questions that
are often collapsed into one: \emph{is this credential authentic?} and
\emph{is this issuer authorized to issue this kind of credential in this
domain?} In decentralized and multi-domain environments, both questions
must be answered.

\subparagraph{7.6.1.3 Multi-Stakeholder Trust
Anchors}\label{multi-stakeholder-trust-anchors}

The model should also support transactions that require validation
against multiple trust anchors. A credential used in a defense and
aerospace workflow, for example, may need to be acceptable both to a
technical standards body that governs material attestations and to a
defense or export-control authority that governs regulatory standing.
The evaluator's local policy determines which trust anchors are required
for a given transaction and whether all, or only a specified subset,
must be satisfied before the artifact can be accepted.

This capability becomes increasingly important in what some emerging
initiatives describe as an Internet of Agents, where autonomous systems
operate across overlapping technical, commercial, and regulatory trust
domains. A single cryptographically valid credential may therefore still
be insufficient unless it can also be situated within the appropriate
governance framework.

\subsubsection{7.7 Delegation and
Attenuation}\label{delegation-and-attenuation}

Delegation allows an authorized entity to grant a subset of its own
authority to another entity. The model enforces a monotonic attenuation
invariant: at every delegation hop, the delegate's authority must be
equal to or narrower than the delegator's.

Permissions can only be narrowed. Constraints can only be tightened.
Constraints cannot be dropped, since omission would widen effective
authority. A delegated credential must therefore retain every parent
constraint, either unchanged or in a strictly attenuated form according
to the semantics of the relevant constraint type.

Delegation should normally remain local to a shared trust and governance
domain. This paper does not recommend cross-trust-boundary delegation as
a default model for inter-organizational agent interaction. Across trust
boundaries, the preferred pattern is independently authorized service
invocation rather than propagation of derived authority. Although
cross-boundary delegation is technically feasible, it creates
substantially greater legal, regulatory, compliance, security, and audit
complexity than independently authorized invocation. It complicates
attribution of responsibility, weakens clarity about which policy regime
governs a given action, increases the risk of unauthorized onward use or
transfer of sensitive authority and data, and creates a form of fate
sharing in which compromise of one trust domain may expose delegated
authority originating elsewhere.

The concern is not merely implementation difficulty. A cryptographic
chain can show where authority originated, but it does not by itself
determine which party controlled the runtime environment, internal
policy logic, operational safeguards, or downstream handling that
produced the action. Cross-boundary delegation therefore creates four
recurring risks: an audit and accountability gap, regulatory and
data-sovereignty friction, compromise-driven fate sharing, and semantic
misalignment between the permission models of independent organizations.
Service invocation avoids much of this complexity because each party
remains independently authorized within its own policy domain.

The appropriate response to dynamic agent spawning is not to rely on
prompting alone, but on cryptographic identity enforcement. In this
model, an external agent is recognized as a valid actor only if it can
present its own cryptographically verifiable subject identity together
with a valid signed authorization credential and proof of possession. A
dynamically spawned sub-agent that lacks its own recognized identity,
valid signature chain, and bound authorization is therefore not treated
as an independently authorized principal by the receiver. From the
perspective of the enforcement engine, it is merely an internal process
of the originating system. Because the receiving system is required to
verify subject binding, issuer trust, audience, and applicable policy
constraints for the presenting actor, such a spawned agent will fail
authorization at the boundary unless it is independently authorized
under an explicitly accepted governance model.

This boundary check is also essential for auditability and
non-repudiation. Each externally acting agent must be attributable to a
distinct cryptographically verifiable identity so that responsibility
for a given action can be reconstructed and defended later. If
dynamically spawned agents were allowed to act across trust boundaries
without their own bound credentials, the resulting audit trail would
become ambiguous: the cryptographic record might show where authority
originated, but not which runtime instance actually exercised it, under
which safeguards, or under whose operational control. Requiring
independently verifiable identity at the receiver therefore protects not
only authorization correctness, but also accountability, evidentiary
integrity, and non-repudiable attribution.

This distinction also helps separate delegation from service invocation.
When one agent calls another as a service, each party may operate under
its own authorization. Delegation, by contrast, creates a chain of
derived authority. Keeping this distinction explicit helps avoid
unverifiable authority propagation in multi-agent systems. For this
reason, the attenuation rules defined here are intended primarily for
local delegation within a shared trust domain, or for exceptional
cross-boundary cases governed by explicit bilateral or profile-level
arrangements.

\begin{center}
\includegraphics[width=0.9\textwidth]{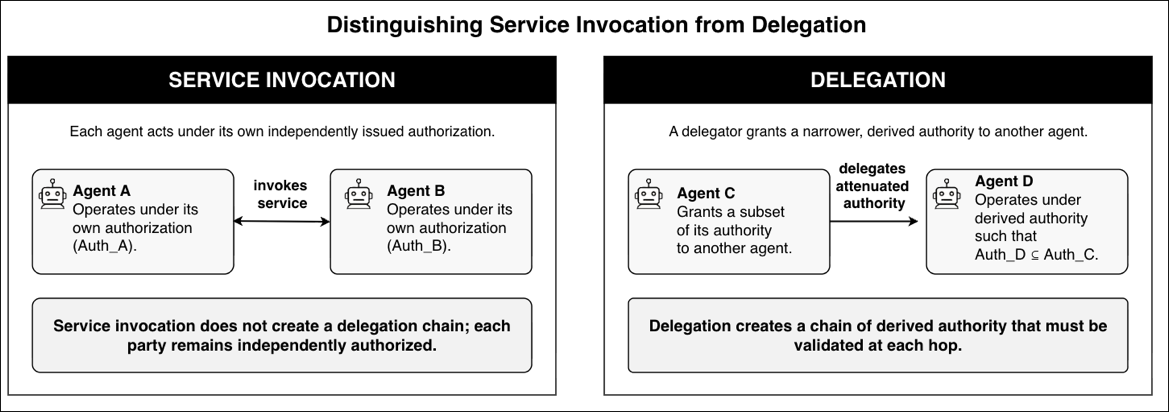}
\end{center}

\emph{Figure 5: This paper distinguishes service invocation from
delegation and recommends service invocation across trust boundaries,
reserving delegation primarily for local trust domains.}

\subsubsection{7.8 Audience Binding}\label{audience-binding}

A credential should be bound not only to its subject but also to its
intended audience. Without audience binding, an agent holding an
authorization credential for one receiver could present that same
credential to another receiver whose trust relationship with the issuer
may be entirely different.

In a JWT profile {[}1{]}, this maps naturally to the \texttt{aud} claim.
In a VC profile, it may be carried in a dedicated audience field within
the authorization payload. The evaluator should reject any credential
whose audience does not include the evaluator itself.

\subsubsection{7.9 Revocation}\label{revocation}

Revocation introduces an online dependency regardless of encoding. A
credential that survives beyond a single request-response exchange
requires a mechanism by which the issuer can signal that a previously
valid credential is no longer active.

The behavioral requirements are straightforward. The issuer must be able
to mark a credential as revoked after issuance. The evaluator must check
revocation status during each evaluation attempt. A revoked credential
must result in denial regardless of whether the embedded constraints
would otherwise pass. Revocation of a parent credential in a delegation
chain must invalidate all downstream delegations derived from it.

In practice, mechanisms such as status lists or comparable revocation
registries can satisfy this requirement. The specific mechanism is a
profile concern. The semantics are not.

\subsubsection{7.10 Signed Audit Trail}\label{signed-audit-trail}

Every evaluation, allow or deny, should produce a cryptographically
protected audit record. The record should include the credential
identifier, agent identity, issuer identity, requested action, requested
resource, relevant request context, each constraint's evaluation result,
the final decision, a timestamp, and the evaluator's signature. Where
the decision is part of a multi-agent workflow, the audit record should
also include the workflow identifier, workflow role, workflow step, and
the set of independently evaluated credentials or attestations relied on
for the final outcome.

This record serves a function analogous to a flight data recorder. It
does not prevent incidents, but it ensures that when an incident occurs,
a disputed settlement, a compliance violation, a suspect supply-chain
attestation, the full chain of evidence is available: who authorized the
agent, what constraints were in effect, what the agent requested, and
why the system allowed or denied it.

\subsubsection{7.11 Canonical Processing
Pipeline}\label{canonical-processing-pipeline}

The following pseudocode illustrates the canonical processing pipeline
for single-credential evaluation. A conformant implementation, given the
same signed authorization, request context, applicable profile and
vocabulary versions, and canonical typing and mapping rules, should
produce the same authorization decision regardless of the enforcement
engine used.

\begin{Shaded}
\begin{Highlighting}[]
\NormalTok{function evaluate(credential, presenterIdentity, requestContext, localPolicy, trustedIssuers):}

\NormalTok{    container = parseContainer(credential)}

\NormalTok{    verificationResult = verifyContainer(}
\NormalTok{        container,}
\NormalTok{        presenterIdentity,}
\NormalTok{        trustedIssuers}
\NormalTok{    )}
\NormalTok{    if verificationResult is not OK:}
\NormalTok{        return verificationResult}

\NormalTok{    payload = extractAuthorizationPayload(container)}
\NormalTok{    if payload is missing agentId, issuerId, permissions, or constraints:}
\NormalTok{        return DENY("credential\_incomplete")}

\NormalTok{    return evaluatePayload(payload, requestContext, localPolicy)}


\NormalTok{function verifyContainer(container, presenterIdentity, trustedIssuers):}

\NormalTok{    if container.signature is invalid:}
\NormalTok{        return DENY("signature\_invalid")}

\NormalTok{    // resolveIssuer includes identifier resolution plus governance validation}
\NormalTok{    issuer = resolveIssuer(container.issuerIdentity)}
\NormalTok{    if issuer not in trustedIssuers:}
\NormalTok{        return DENY("issuer\_untrusted")}

\NormalTok{    if governanceRegistry does not recognize issuer for this credential type / profile:}
\NormalTok{        return DENY("issuer\_not\_vetted")}

\NormalTok{    if container.audience is defined and evaluatorIdentity not in container.audience:}
\NormalTok{        return DENY("audience\_mismatch")}

\NormalTok{    if presenter does not prove possession of container.subjectIdentity:}
\NormalTok{        return DENY("proof\_of\_possession\_failed")}

\NormalTok{    if container.subjectIdentity != presenterIdentity:}
\NormalTok{        return DENY("subject\_binding\_mismatch")}

\NormalTok{    if container is expired:}
\NormalTok{        return DENY("credential\_expired")}

\NormalTok{    if container is revoked:}
\NormalTok{        return DENY("credential\_revoked")}

\NormalTok{    return OK}


\NormalTok{function evaluatePayload(payload, requestContext, localPolicy):}

\NormalTok{    if requestContext.action not in payload.permissions:}
\NormalTok{        return DENY("permission\_denied")}

\NormalTok{    for each constraint in payload.constraints:}
\NormalTok{        if constraint.type is unknown:}
\NormalTok{            return DENY("constraint\_unknown")}
\NormalTok{        if requestContext[constraint.field] is absent:}
\NormalTok{            return DENY("context\_field\_missing")}
\NormalTok{        if evaluateConstraint(constraint, requestContext[constraint.field]) is FAIL:}
\NormalTok{            return DENY("constraint\_failed")}

\NormalTok{    for each localConstraint in localPolicy.constraints:}
\NormalTok{        if requestContext[localConstraint.field] is absent:}
\NormalTok{            return DENY("context\_field\_missing")}
\NormalTok{        if evaluateConstraint(localConstraint, requestContext[localConstraint.field]) is FAIL:}
\NormalTok{            return DENY("local\_policy\_denied")}

\NormalTok{    return ALLOW}
\end{Highlighting}
\end{Shaded}

For local delegation chains, or for exceptional cross-boundary cases
governed by explicit bilateral or profile-level arrangements, the
pipeline is extended with a chain verification phase before terminal
payload evaluation. Each link is checked for signature validity,
revocation status, issuer-subject continuity, permission narrowing, and
constraint attenuation. Any widening of authority, including omission of
parent constraints, invalidates the chain.

\begin{Shaded}
\begin{Highlighting}[]
\NormalTok{function evaluateDelegationChain(credentials[], presenterIdentity, requestContext, localPolicy, trustedIssuers, maxChainDepth):}

\NormalTok{    // Intended for local delegation chains or explicitly governed cross{-}boundary exceptions}
\NormalTok{    if length(credentials) \textgreater{} maxChainDepth:}
\NormalTok{        return DENY("delegation\_depth\_exceeded")}

\NormalTok{    containers = []}
\NormalTok{    payloads = []}

\NormalTok{    for i = 0 to length(credentials) {-} 1:}
\NormalTok{        container = parseContainer(credentials[i])}
\NormalTok{        payload = extractAuthorizationPayload(container)}

\NormalTok{        if payload is missing agentId, issuerId, permissions, or constraints:}
\NormalTok{            return DENY("credential\_incomplete")}

\NormalTok{        containers.append(container)}
\NormalTok{        payloads.append(payload)}

\NormalTok{    for i = 0 to length(containers) {-} 1:}
\NormalTok{        expectedPresenter =}
\NormalTok{            presenterIdentity if i == length(containers) {-} 1}
\NormalTok{            else containers[i].subjectIdentity}

\NormalTok{        verificationResult = verifyContainer(}
\NormalTok{            containers[i],}
\NormalTok{            expectedPresenter,}
\NormalTok{            trustedIssuers}
\NormalTok{        )}
\NormalTok{        if verificationResult is not OK:}
\NormalTok{            return verificationResult}

\NormalTok{        if i \textgreater{} 0:}
\NormalTok{            if containers[i].issuerIdentity != containers[i {-} 1].subjectIdentity:}
\NormalTok{                return DENY("delegation\_chain\_broken")}

\NormalTok{            if payloads[i].permissions not subset of payloads[i {-} 1].permissions:}
\NormalTok{                return DENY("delegation\_widened")}

\NormalTok{            for each constraint in payloads[i].constraints:}
\NormalTok{                if attenuation(constraint, correspondingParentConstraint) is invalid:}
\NormalTok{                    return DENY("delegation\_widened")}

\NormalTok{            for each parentConstraint in payloads[i {-} 1].constraints not represented in payloads[i].constraints:}
\NormalTok{                return DENY("delegation\_widened")}

\NormalTok{    return evaluatePayload(payloads[last], requestContext, localPolicy)}
\end{Highlighting}
\end{Shaded}

\begin{center}
\includegraphics[width=0.9\textwidth]{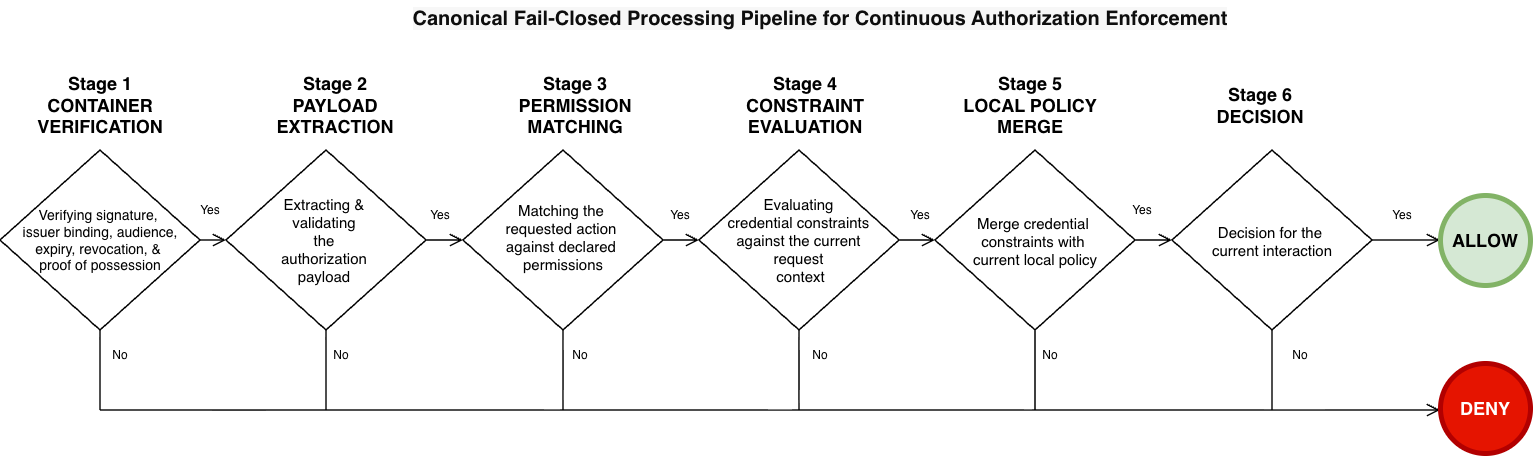}
\end{center}

\emph{Figure 6: Authorization is re-evaluated by the receiving system at
each interaction; any failed stage results in immediate denial
(fail-closed).}

\paragraph{7.11.1 Worked Trace: Insurance Settlement
Negotiation}\label{worked-trace-insurance-settlement-negotiation}

To illustrate the processing pipeline end to end, this section walks
through a single authorization decision from the insurance claims use
case described in Section 4.

\textbf{Credential.} MegaInsure's IAM system issues a JWT/JWS credential
to its negotiator agent, \texttt{agent:megainsure:negotiator-7}. The
signed authorization payload contains:

{\def\LTcaptype{none} 
\begin{longtable}[]{@{}ll@{}}
\toprule\noalign{}
Component & Value \\
\midrule\noalign{}
\endhead
\bottomrule\noalign{}
\endlastfoot
Agent Identity & \texttt{agent:megainsure:negotiator-7} \\
Issuer Identity & \texttt{iss:megainsure:claims-authority} \\
Permissions & \texttt{claim.settle} \\
Constraints & see below \\
\end{longtable}
}

The constraint set carried in the credential is:

{\def\LTcaptype{none} 
\footnotesize
\begin{longtable}[]{@{}
  >{\raggedright\arraybackslash}p{(\linewidth - 8\tabcolsep) * \real{0.0400}}
  >{\raggedright\arraybackslash}p{(\linewidth - 8\tabcolsep) * \real{0.2600}}
  >{\raggedright\arraybackslash}p{(\linewidth - 8\tabcolsep) * \real{0.1600}}
  >{\raggedright\arraybackslash}p{(\linewidth - 8\tabcolsep) * \real{0.3400}}
  >{\raggedright\arraybackslash}p{(\linewidth - 8\tabcolsep) * \real{0.2000}}@{}}
\toprule\noalign{}
\begin{minipage}[b]{\linewidth}\raggedright
\#
\end{minipage} & \begin{minipage}[b]{\linewidth}\raggedright
Type
\end{minipage} & \begin{minipage}[b]{\linewidth}\raggedright
Field
\end{minipage} & \begin{minipage}[b]{\linewidth}\raggedright
Operator / Values
\end{minipage} & \begin{minipage}[b]{\linewidth}\raggedright
Purpose
\end{minipage} \\
\midrule\noalign{}
\endhead
\bottomrule\noalign{}
\endlastfoot
C1 & \texttt{Temporal\-Window\-Constraint} & \texttt{core.request\_time} &
\texttt{valid\_from}: \texttt{2026-04-18T00:00:00Z},
\texttt{valid\_until}: \texttt{2026-04-18T23:59:59Z} & 24-hour
authorization window \\
C2 & \texttt{Numeric\-Limit\-Constraint} & \texttt{core.amount} &
\texttt{lte}, \texttt{USD\ 5,000} & Settlement ceiling \\
C3 & \texttt{Numeric\-Limit\-Constraint} & \texttt{core.amount} &
\texttt{gte}, \texttt{USD\ 500} & Settlement floor \\
C4 & \texttt{Enumerated\-List\-Constraint} & \texttt{insurance.claim\_type}
& \texttt{\{auto\_collision,\ auto\_comprehensive\}} & Claim-type
restriction \\
\end{longtable}
}

\textbf{Request context.} The negotiator agent presents the credential
to BodyShopCo's enforcement engine and requests \texttt{claim.settle}
for \texttt{core.resource\_id=claims/auto/CLM-90421}, with
\texttt{core.amount=USD\ 3,200},
\texttt{insurance.claim\_type=auto\_collision},
\texttt{core.workflow\_id=CLM-90421}, and
\texttt{core.request\_time=2026-04-18T14:32:00Z}.

\textbf{Pipeline execution.}

{\def\LTcaptype{none} 
\small
\begin{longtable}[]{@{}
  >{\raggedright\arraybackslash}p{(\linewidth - 6\tabcolsep) * \real{0.0600}}
  >{\raggedright\arraybackslash}p{(\linewidth - 6\tabcolsep) * \real{0.2600}}
  >{\raggedright\arraybackslash}p{(\linewidth - 6\tabcolsep) * \real{0.5400}}
  >{\raggedright\arraybackslash}p{(\linewidth - 6\tabcolsep) * \real{0.1400}}@{}}
\toprule\noalign{}
\begin{minipage}[b]{\linewidth}\raggedright
Step
\end{minipage} & \begin{minipage}[b]{\linewidth}\raggedright
Function
\end{minipage} & \begin{minipage}[b]{\linewidth}\raggedright
Check
\end{minipage} & \begin{minipage}[b]{\linewidth}\raggedright
Result
\end{minipage} \\
\midrule\noalign{}
\endhead
\bottomrule\noalign{}
\endlastfoot
1 & \texttt{parseContainer} & JWT structure is valid & OK \\
2 & \texttt{verifyContainer}: signature & JWS signature verifies against
MegaInsure's published key & OK \\
3 & \texttt{verifyContainer}: issuer &
\texttt{iss:megainsure:claims-authority} is in BodyShopCo's trusted
issuer set & OK \\
4 & \texttt{verifyContainer}: audience & \texttt{aud} includes
\texttt{svc:bodyshopco:claims-api} & OK \\
5 & \texttt{verifyContainer}: proof of possession & Agent proves
possession of the subject key & OK \\
6 & \texttt{verifyContainer}: expiry and revocation & Credential is not
expired and not revoked & OK \\
7 & \texttt{evaluatePayload}: permission & \texttt{claim.settle} is in
the declared permission set & OK \\
8 & \texttt{evaluatePayload}: C1 & \texttt{2026-04-18T14:32:00Z} is
within the authorized time window & PASS \\
9 & \texttt{evaluatePayload}: C2 &
\texttt{USD\ 3,200\ \textless{}=\ USD\ 5,000} & PASS \\
10 & \texttt{evaluatePayload}: C3 &
\texttt{USD\ 3,200\ \textgreater{}=\ USD\ 500} & PASS \\
11 & \texttt{evaluatePayload}: C4 & \texttt{auto\_collision} is in
\texttt{\{auto\_collision,\ auto\_comprehensive\}} & PASS \\
12 & Local policy & BodyShopCo requires \texttt{core.workflow\_id} to be
present & PASS: \texttt{CLM-90421} \\
\textbf{13} & \textbf{Decision} & \textbf{All credential constraints and
local policy checks pass} & \textbf{ALLOW} \\
\end{longtable}
}

\textbf{Denial variant.} If the negotiator instead requests
\texttt{core.amount=USD\ 7,500}, step 9 evaluates
\texttt{USD\ 7,500\ \textless{}=\ USD\ 5,000} as false. The pipeline
returns \texttt{DENY("constraint\_failed")} for C2. The audit record
captures the credential digest, denial reason, failed constraint
identifier, and request context snapshot, allowing MegaInsure and
BodyShopCo to determine whether the agent exceeded its authority or
whether the credential scope was insufficient for the claim.

\subsubsection{7.12 Semantic Interoperability and
Resolution}\label{semantic-interoperability-and-resolution}

The model defines the grammar of authorization: the structure and
algebra of constraints. Portability across heterogeneous receivers,
however, also requires governed semantic alignment. Literal string
matching for field identifiers is too brittle: an issuer may sign a
constraint against one identifier while a receiver organizes the
relevant request context under another. In that case, evaluation will
correctly fail closed even though the signed authority is otherwise
valid. Portable authorization therefore requires a semantic resolution
layer that separates the stable horizontal core of the model from
governed vertical industry vocabularies and mapping profiles.

\paragraph{7.12.1 Hub-and-Spoke
Architecture}\label{hub-and-spoke-architecture}

To preserve decision consistency without requiring uniform internal
schemas, the model adopts a hub-and-spoke architecture for semantic
identification. The hub consists of a governed Minimum Viable Vocabulary
for shared authorization primitives, while the spokes consist of
versioned industry profiles that extend that core for domain-specific
use.

The \textbf{horizontal core} defines a set of reserved,
industry-agnostic semantic identifiers that every conformant enforcement
engine must recognize and interpret consistently. These identifiers form
the common semantic substrate for the core constraint algebra.

The \textbf{vertical profiles} define domain-specific vocabularies for
industries such as insurance or aerospace. These profiles introduce
specialized fields, actions, and aliases, but do so under explicit
governance and versioning. They are discoverable through the trust and
governance frameworks described in Section 7.6.1 and are intended to
extend the core rather than fragment it.

This architecture allows the semantic model to remain stable across
domains while giving industries room to express their own specialized
concepts without sacrificing portability.

\paragraph{7.12.2 Minimum Viable
Vocabulary}\label{minimum-viable-vocabulary}

To prevent each deployment from becoming its own integration silo, the
model defines a \textbf{Minimum Viable Vocabulary} consisting of shared
authorization primitives that every conformant enforcement engine must
recognize without requiring live external resolution at evaluation time.
These primitives are represented as reserved semantic identifiers so
that the evaluator can remain local, deterministic, and fail-closed.

For the purposes of this model, the horizontal core vocabulary is the
enumerated set below. The table defines the reserved identifier,
canonical type, status, and intended meaning. \texttt{Required} means
that every conformant evaluator must recognize the identifier and its
type. \texttt{Conditional} means that the identifier is not required in
every request, but if a credential, constraint, or profile references
it, the evaluator must resolve it through a trusted mapping profile or
fail closed. \texttt{Advanced} means that the identifier is reserved for
implementations that claim support for stateful governance profiles.

{\def\LTcaptype{none} 
\begin{longtable}[]{@{}
  >{\raggedright\arraybackslash}p{(\linewidth - 6\tabcolsep) * \real{0.2500}}
  >{\raggedright\arraybackslash}p{(\linewidth - 6\tabcolsep) * \real{0.2500}}
  >{\raggedright\arraybackslash}p{(\linewidth - 6\tabcolsep) * \real{0.2500}}
  >{\raggedright\arraybackslash}p{(\linewidth - 6\tabcolsep) * \real{0.2500}}@{}}
\toprule\noalign{}
\begin{minipage}[b]{\linewidth}\raggedright
Identifier
\end{minipage} & \begin{minipage}[b]{\linewidth}\raggedright
Type
\end{minipage} & \begin{minipage}[b]{\linewidth}\raggedright
Status
\end{minipage} & \begin{minipage}[b]{\linewidth}\raggedright
Description
\end{minipage} \\
\midrule\noalign{}
\endhead
\bottomrule\noalign{}
\endlastfoot
\texttt{core.issuer\_id} & String identifier & Required & Entity that
issued and signed the authorization. \\
\texttt{core.subject\_id} & String identifier & Required & Agent or
principal to whom the authorization applies. \\
\texttt{core.presenter\_id} & String identifier & Required & Runtime
actor presenting the credential at the receiver boundary. \\
\texttt{core.audience\_id} & String identifier & Required & Intended
receiver or receiver class for audience binding. \\
\texttt{core.permission} & String identifier & Required & Declared
action or permission carried by the authorization payload. \\
\texttt{core.valid\_from} & Timestamp & Required & Earliest time at
which the authorization may be accepted. \\
\texttt{core.valid\_until} & Timestamp & Required & Latest time at which
the authorization may be accepted. \\
\texttt{core.request\_time} & Timestamp & Required & Time of the
action-bearing request, normalized for temporal evaluation. \\
\texttt{core.delegator\_id} & String identifier & Conditional & Prior
authority in a delegation chain. Required when delegated authority is
evaluated. \\
\texttt{core.recipient\_id} & String identifier & Conditional &
Counterparty, payee, recipient system, or recipient role for recipient
restrictions. \\
\texttt{core.action} & String identifier & Conditional & Requested
operation when a receiver distinguishes action from permission
namespace. \\
\texttt{core.resource\_id} & String identifier & Conditional & Specific
object, record, document, component, or resource being acted on. \\
\texttt{core.resource\_type} & String identifier & Conditional &
Category of resource, such as claim, invoice, part, shipment, or
attestation. \\
\texttt{core.amount} & Decimal number & Conditional & Monetary or
quantitative value for the current action. \\
\texttt{core.currency\_code} & String code & Conditional & Currency
associated with \texttt{core.amount}, where monetary limits are used. \\
\texttt{core.quantity} & Decimal number & Conditional & Non-monetary
measured amount, such as units, weight, volume, or countable
quantity. \\
\texttt{core.count} & Integer & Conditional & Number of items, records,
disclosures, attempts, or actions in the current request. \\
\texttt{core.total\_budget} & Decimal number & Conditional & Maximum
aggregate value authorized by the credential or applicable profile. \\
\texttt{core.geo\_region} & String identifier & Conditional & Geographic
region, jurisdiction, or data-sovereignty boundary used for
location-sensitive constraints. \\
\texttt{core.ip\_address} & IP address string & Conditional & Network
source address when network-origin restrictions are part of the
governing profile. \\
\texttt{core.request\_id} & String identifier & Conditional &
Receiver-local or cross-domain request identifier used for audit
correlation. \\
\texttt{core.workflow\_id} & String identifier & Conditional &
Correlation identifier for a multi-agent or multi-step workflow. \\
\texttt{core.workflow\_role} & String identifier & Conditional & Role
played by the presenting agent within a collaborative workflow. \\
\mbox{\texttt{core.workflow\_step\_id}} & String identifier & Conditional &
Specific workflow step, approval stage, or evidence contribution being
authorized. \\
\texttt{core.state\_authority\_pointer} & URI or string identifier &
Advanced & Authoritative state source, registry, ledger, or voucher
authority for cumulative governance. \\
\texttt{core.state\_sequence} & Integer & Advanced & Monotonic sequence
value used by verifiable state proofs to prevent rollback or replay. \\
\texttt{core.state\_timestamp} & Timestamp & Advanced & Signed state
observation time used to evaluate freshness within the applicable drift
envelope. \\
\end{longtable}
}

Runtime enforcement remains local and deterministic. Semantic agreement
is established in advance through governed vocabularies and mapping
profiles rather than through live runtime dereferencing.

\paragraph{7.12.3 Vertical Industry Profiles and
Aliasing}\label{vertical-industry-profiles-and-aliasing}

Domain vocabularies extend the horizontal core but do not replace it. A
conformant implementation uses governed semantic aliasing to map local
request structures to shared semantic identifiers defined either in the
core or in an applicable governed industry profile.

When a credential is presented, the enforcement engine must determine
how the signed semantic identifiers in the credential map to the
receiver's local request structure. It does so by consulting a
machine-readable mapping profile that defines approved aliases between
local field names and the shared identifiers defined by the core
vocabulary or an applicable governed profile. If a local field is
explicitly declared as an alias of the signed identifier, the engine may
resolve the names and evaluate the constraint against the corresponding
local field. The purpose of this step is not to reinterpret the
authorization, but to reconcile naming differences while preserving the
semantics of the signed policy.

This model achieves portability without requiring every receiver to
adopt identical internal schema names. Just as importantly, it separates
\textbf{semantic agreement} from \textbf{runtime enforcement}: mappings
are established in advance through governed profiles, while runtime
evaluation operates only on pre-resolved identifiers and validated
aliases.

\paragraph{7.12.4 Type-Safe Coercion and Risk
Mitigation}\label{type-safe-coercion-and-risk-mitigation}

Moving from literal string matching to governed semantic mapping reduces
one class of interoperability failure, but it also creates a need for
strong guardrails. To preserve safety, the enforcement engine must apply
type-safe coercion rules. Any mapping that produces an ambiguous type
conversion or an uncertain semantic match must result in denial.

A \texttt{Numeric\-Limit\-Constraint} can be mapped to a local field only if
that field is explicitly typed as numeric in the applicable mapping
profile. A categorical constraint must not be coerced into a monetary
field merely because the local schema is underspecified or loosely
typed. If a mapping profile is missing, unreachable, stale beyond its
permitted governance window, or contains conflicting aliases for the
same field, the evaluator must fail closed.

Profiles must declare their supported vocabularies explicitly so that
semantic compatibility between an agent and a service provider can be
validated before an action-bearing interaction begins. This preserves
the fail-closed property of the broader model while allowing
authorization semantics to remain portable across heterogeneous systems.

\paragraph{7.12.5 Semantic Field Resolution as a Diagnostic
Gate}\label{semantic-field-resolution-as-a-diagnostic-gate}

Semantic resolution serves as a diagnostic gate between the portable
authorization credential and the receiver's local execution context. Its
function is to determine whether a signed semantic identifier can be
mapped safely and unambiguously to a local field under the applicable
governed profile. The model does not attempt to automate all semantic
recovery. Instead, when resolution fails, the evaluator fails closed and
returns a typed denial reason. That denial reason may be used by
external orchestration, human review, or out-of-band profile
reconciliation mechanisms, but any such recovery remains outside the
normative evaluation pipeline.

The following pseudocode illustrates the canonical fail-closed procedure
for resolving signed semantic identifiers to local request fields prior
to constraint evaluation.

\begin{Shaded}
\begin{Highlighting}[]
\NormalTok{function resolveSemanticField(constraint, requestContext, mappingProfile, applicableProfileVocabulary):}

\NormalTok{    if mappingProfile is missing:}
\NormalTok{        return DENY("mapping\_profile\_missing")}

\NormalTok{    if mappingProfile is stale beyond its permitted governance window or untrusted:}
\NormalTok{        return DENY("mapping\_profile\_invalid")}

\NormalTok{    signedIdentifier = constraint.field}

\NormalTok{    if signedIdentifier is in coreVocabulary:}
\NormalTok{        expectedType = coreVocabulary[signedIdentifier].type}
\NormalTok{    else if signedIdentifier is in applicableProfileVocabulary:}
\NormalTok{        expectedType = applicableProfileVocabulary[signedIdentifier].type}
\NormalTok{    else:}
\NormalTok{        return DENY("semantic\_identifier\_unknown")}

\NormalTok{    if mappingProfile has conflicting aliases for signedIdentifier:}
\NormalTok{        return DENY("semantic\_alias\_conflict")}

\NormalTok{    localField = mappingProfile.resolveAlias(signedIdentifier)}

\NormalTok{    if localField is missing:}
\NormalTok{        return DENY("semantic\_alias\_missing")}

\NormalTok{    if mappingProfile.typeOf(localField) != expectedType:}
\NormalTok{        return DENY("semantic\_type\_mismatch")}

\NormalTok{    if requestContext[localField] is absent:}
\NormalTok{        return DENY("context\_field\_missing")}

\NormalTok{    return OK(requestContext[localField])}
\end{Highlighting}
\end{Shaded}

\paragraph{7.12.5.1 Constraint Enforcement After
Resolution}\label{constraint-enforcement-after-resolution}

Once semantic resolution succeeds, constraint enforcement proceeds
against the resolved field without altering the meaning of the signed
authorization.

\begin{Shaded}
\begin{Highlighting}[]
\NormalTok{resolutionResult = resolveSemanticField(}
\NormalTok{    constraint,}
\NormalTok{    requestContext,}
\NormalTok{    mappingProfile,}
\NormalTok{    applicableProfileVocabulary}
\NormalTok{)}

\NormalTok{if resolutionResult is not OK:}
\NormalTok{    return resolutionResult}

\NormalTok{resolvedValue = resolutionResult.value}

\NormalTok{if evaluateConstraint(constraint, resolvedValue) is FAIL:}
\NormalTok{    return DENY("constraint\_failed")}
\end{Highlighting}
\end{Shaded}

\paragraph{7.12.5.2 Determinism Invariant and External
Recovery}\label{determinism-invariant-and-external-recovery}

The semantic resolution layer must preserve the determinism of the
broader authorization model. For any fixed signed authorization
credential \texttt{C}, request context \texttt{R}, applicable profile
and vocabulary versions \texttt{V}, and governed mapping profile
\texttt{M}, a conformant evaluator must produce a single authorization
decision \emph{D} \(\in\) \{ALLOW, DENY\}. The evaluator may
additionally emit typed diagnostic metadata, but such metadata does not
alter the decision domain.

This constraint is essential. Any mechanism that leaves the
authorization outcome dependent on interactive negotiation,
probabilistic recovery, or implementation-specific semantic guesswork
would violate the determinism invariant on which portability and
standardization depend. Semantic resolution is therefore not an
open-ended attempt to recover meaning from arbitrary local schemas. It
is a governed validation step that either resolves the signed identifier
safely and unambiguously, or fails closed.

Typed denial reasons remain important even within this binary model. A
denial caused by \texttt{semantic\_alias\_missing},
\texttt{mapping\_profile\_invalid}, \texttt{semantic\_type\_mismatch},
or \mbox{\texttt{context\_field\_missing}} may be used by surrounding systems
as a diagnostic signal for metadata refresh, bilateral profile
alignment, human review, or other out-of-band remediation. Such recovery
mechanisms may be operationally useful, but they remain external to the
normative evaluation pipeline. The role of the standard is to ensure
that every engagement attempt yields a verifiable, deterministic, and
diagnosable outcome, and that autonomous execution never proceeds under
conditions of unresolved semantic ambiguity.

\subsubsection{7.13 Pre-flight Discovery
Protocol}\label{pre-flight-discovery-protocol}

Portable authorization is not operationally useful if semantic
compatibility is discovered only at the moment of enforcement. In that
model, otherwise valid integrations become brittle: credentials fail
because of vocabulary mismatch, unsupported profiles, incompatible trust
anchors, or missing context fields that could have been identified
earlier. The purpose of pre-flight discovery is therefore not to
guarantee that every workflow will succeed, but to reduce integration
brittleness by turning semantic incompatibility into an explicit,
deterministic, and diagnosable admission outcome before an
action-bearing request is attempted.

Under this model, the receiver publishes a signed, versioned, and
cacheable governance contract describing the vocabularies, profiles,
trust anchors, and required context fields it accepts. A sender may
resolve its credential fields against this published contract before
presentation in order to determine likely compatibility, but that step
is advisory rather than authoritative. Authoritative interpretation
remains entirely with the receiver's enforcement engine, which performs
the same semantic resolution and policy evaluation at admission time. If
the sender determines pre-flight that its credential is incompatible,
the workflow fails early, or is rerouted before runtime effort is
wasted. If the sender incorrectly concludes that it is compatible, the
receiver still remains authoritative and denies the request. In that
case, the failure reflects contract drift, stale metadata, or a
non-conformant sender implementation, not a failure of receiver-side
enforcement.

\paragraph{7.13.1 Discovery Endpoint}\label{discovery-endpoint}

A conformant service provider should expose a standardized discovery
endpoint, such as \texttt{/.well-known/agent-governance} {[}8{]}, that
publishes this governance contract as a machine-readable manifest. The
manifest is a public interoperability declaration, not a disclosure of
private policy logic. At a minimum, it declares the supported
vocabularies and profile versions, accepted trust anchors or trust
registries, required request-context fields, and the version and
freshness metadata needed for safe caching and validation.

To prevent tampering or misbinding, the manifest must itself be
cryptographically signed by the receiver, or otherwise
integrity-protected in a manner verifiable against the same trust
infrastructure used to authenticate the receiver. Because the manifest
is static, version-controlled metadata rather than per-request
negotiation state, it can be cached safely within its declared validity
window. Senders may consult it pre-flight to estimate compatibility,
while the receiver's enforcement engine, or an authoritative local
equivalent of the same manifest, remains solely responsible for final
semantic resolution and authorization decisions at admission time. The
manifest therefore reduces ambiguity without shifting decision authority
away from the receiver.

\paragraph{7.13.2 Discovery Sequence}\label{discovery-sequence}

The discovery process follows a structured sequence designed to reduce
operational friction while preserving receiver authority. First, the
sender retrieves the receiver's published governance manifest, either
directly or from a cached copy within its declared validity window.
Second, the sender compares that manifest against its own supported
profiles, trust anchors, available credentials, and required context
fields in order to determine likely compatibility. Third, if a
compatible path exists, the sender selects the credential it intends to
present and, where appropriate, attenuates its authority locally so that
the presented credential remains narrowly scoped and consistent with the
receiver's published interoperability contract.

The important design point is that this process occurs before execution
and remains advisory on the sender side. It is a pre-flight
compatibility and governance check, not a per-request negotiation loop
and not a substitute for receiver-side enforcement.

\paragraph{7.13.3 Privacy and Security
Safeguards}\label{privacy-and-security-safeguards}

Pre-flight discovery also serves a privacy and security function. By
consulting the receiver's published governance contract in advance, a
sender can avoid over-sharing or presenting sensitive credentials to an
endpoint that does not advertise a compatible trust or semantic
framework.

The manifest therefore advertises only public interoperability
requirements. It does not disclose the receiver's private internal
policies, local thresholds, or proprietary enforcement logic. Transport
protection such as TLS is helpful but not sufficient on its own; the
manifest must be cryptographically bound to the receiver's identity so
that the sender can verify that the published contract actually
originated from the intended service provider. If the sender determines
pre-flight that it cannot meet the receiver's published trust, profile,
or semantic requirements, the workflow can fail early or be rerouted
before any sensitive authorization credential is presented.

\subsubsection{7.14 Integration with Trust
Registries}\label{integration-with-trust-registries}

The discovery manifest situates an interaction within the broader
governance framework described earlier in Section 7.6.1. By referencing
specific trust anchors, registries, and governed profiles, the receiver
publishes the trust and semantic domain within which presented
credentials will be interpreted.

This supports mutual compatibility checking without weakening receiver
authority. A sender may determine whether it is likely to be compatible
with the receiver's published governance contract before initiating an
action-bearing exchange, while the receiver's enforcement engine remains
solely responsible for final interpretation and admission decisions. The
same mechanism also supports dynamic governance: when an industry
registry changes the standing of an issuer, updates accepted credential
classes, or revises profile support, those changes can propagate into
future interactions through the published manifest without altering the
core evaluation semantics.

A typical governance manifest identifies the receiver, declares the
supported profile versions, specifies which trust registries are
accepted for which credential classes, and enumerates the semantic
context fields required for authorization evaluation. It does not
replace trust registries or the portable authorization credential
itself; rather, it publishes the receiver's interoperability contract in
a form that allows compatibility to be assessed in advance and enforced
consistently at admission time.

\paragraph{7.14.1 Governance of Governance
Artifacts}\label{governance-of-governance-artifacts}

Governed vocabularies, trust registries, and mapping profiles are not
self-authorizing. A conformant ecosystem must treat them as governed
artifacts with clear stewardship, versioning, auditability, and
transition rules. The model does not require a single global authority.
A steward may be an industry consortium, regulator, standards body,
enterprise federation, or bilateral governance group. What matters is
that the receiver can identify which steward it trusts for a given
profile, which public keys or trust anchors authenticate that steward,
and which signed artifact versions are accepted at enforcement time.

Profile conflicts are resolved by explicit selection, not by runtime
negotiation or semantic guessing. If two industry profiles define
incompatible meanings for the same domain concept, they must remain
distinguishable by profile identifier, namespace, and version. A
receiver may choose one, accept both under separate namespaces, or
reject the interaction if the credential depends on an unsupported or
ambiguous profile. The enforcement engine must not merge conflicting
profile semantics on its own.

Governance artifacts should be released as immutable, signed versions
with machine-readable metadata: profile identifier, version, effective
date, expiration or review date, issuing steward, artifact digest,
predecessor or successor relationship, and compatibility notes. Minor
updates may add aliases or clarify descriptions without changing
evaluation meaning. Breaking changes should require a new major profile
version and a defined migration window. Receiver manifests should pin
the accepted versions or version ranges so that both sender-side
pre-flight checks and receiver-side enforcement operate against the same
published contract.

During governance transitions, such as trust-anchor rotation, registry
operator change, or profile deprecation, the safe default is overlap
followed by fail-closed. The outgoing and incoming authorities may both
be accepted only during an explicitly declared transition window and
only if the receiver's governance manifest lists both. Once the
transition window expires, credentials, profiles, mapping profiles, or
state authorities anchored only in the deprecated governance path must
be rejected. Audit records should capture the registry version, profile
version, manifest digest, and trust anchor used for each decision so
that later disputes can be reconstructed from the exact governance state
in force at the time of evaluation.

\begin{center}
\includegraphics[width=0.9\textwidth]{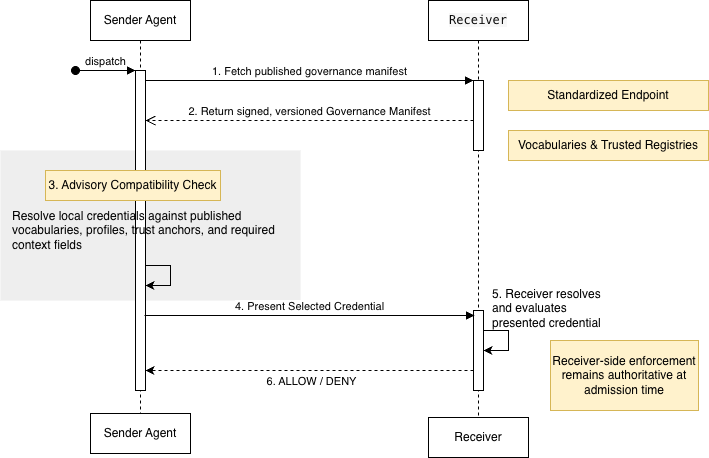}
\end{center}

\emph{Figure 7: Pre-flight discovery uses a published, signed governance
manifest to support advisory compatibility checking before credential
presentation, while preserving authoritative receiver-side enforcement
at admission time.}

\subsubsection{7.15 Illustrative Reference
Architecture}\label{illustrative-reference-architecture}

The preceding sections define the normative semantic and governance
model. The figure below illustrates one possible reference architecture
for implementing published governance contracts, sender-side
compatibility checking, and receiver-side semantic resolution and
enforcement. The figure is illustrative rather than normative:
conformant implementations may realize the same model using different
protocol bindings, trust frameworks, and enforcement components.

\begin{center}
\includegraphics[width=0.9\textwidth]{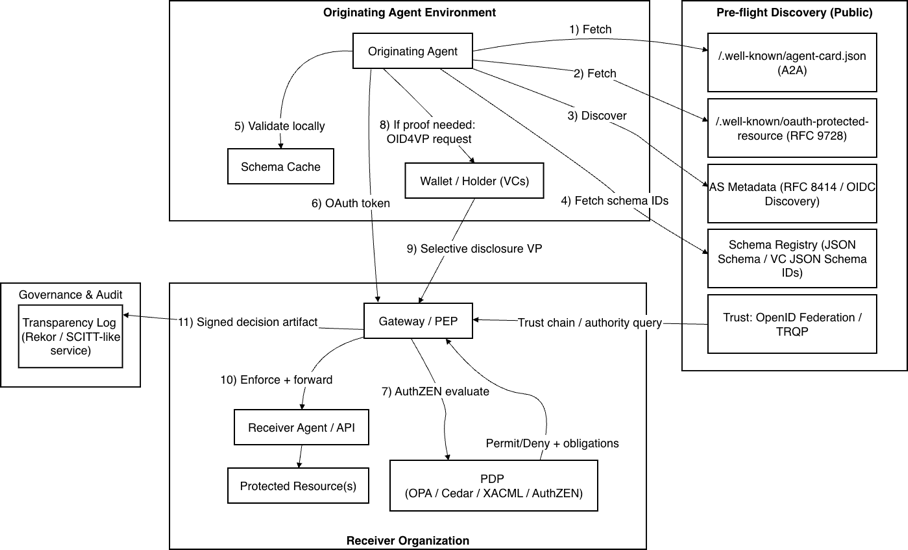}
\end{center}

\emph{Figure 8: Illustrative reference architecture for published
governance contracts, sender-side compatibility checking, and
receiver-side semantic resolution and enforcement.}

\subsubsection{7.16 Threat Model and Security
Considerations}\label{threat-model-and-security-considerations}

The model assumes that standard cryptographic primitives remain secure,
that trusted issuers protect their signing keys, and that receivers
operate their enforcement engines within their own trusted computing
boundary. The attacker may observe, replay, relay, modify, or present
authorization credentials; attempt to exploit semantic mismatches;
poison or replay governance metadata; impersonate state authorities; or
present stale cumulative-state evidence. The attacker is not assumed to
be able to forge valid signatures from uncompromised issuers or break
the underlying cryptographic algorithms.

{\def\LTcaptype{none} 
\begin{longtable}[]{@{}
  >{\raggedright\arraybackslash}p{(\linewidth - 6\tabcolsep) * \real{0.2500}}
  >{\raggedright\arraybackslash}p{(\linewidth - 6\tabcolsep) * \real{0.2500}}
  >{\raggedright\arraybackslash}p{(\linewidth - 6\tabcolsep) * \real{0.2500}}
  >{\raggedright\arraybackslash}p{(\linewidth - 6\tabcolsep) * \real{0.2500}}@{}}
\toprule\noalign{}
\begin{minipage}[b]{\linewidth}\raggedright
\textbf{Threat}
\end{minipage} & \begin{minipage}[b]{\linewidth}\raggedright
\textbf{Attack Pattern}
\end{minipage} & \begin{minipage}[b]{\linewidth}\raggedright
\textbf{Mitigation in the Model}
\end{minipage} & \begin{minipage}[b]{\linewidth}\raggedright
\textbf{Residual Risk}
\end{minipage} \\
\midrule\noalign{}
\endhead
\bottomrule\noalign{}
\endlastfoot
Credential forgery & Attacker fabricates or modifies an authorization
credential & Signature verification, issuer trust validation, proof of
possession & Compromised issuer keys remain a governance and revocation
problem \\
Credential relay & Attacker relays a valid credential to a different
receiver or context & Audience binding, proof of possession, subject
binding, receiver-side evaluation & Stronger profiles may add nonce,
channel, or transaction binding \\
Delegation widening & Delegate attempts to gain broader authority than
parent & Attenuation rules, most-restrictive-wins, denial if parent
constraints are omitted & Incorrect implementation of attenuation
remains a conformance risk \\
Semantic alias poisoning & Mapping profile maps a signed identifier to
the wrong local field & Governed mapping profiles, type-safe coercion,
alias conflict detection, fail-closed behavior & Governance failure in
profile publication or approval remains possible \\
Stale or compromised mapping profile & Receiver uses outdated or
tampered semantic metadata & Signed manifests, versioning, freshness
windows, trust-registry validation & Incorrect cache policy may cause
avoidable denials or stale interpretation \\
Discovery manifest tampering & Attacker publishes false compatibility or
trust metadata & Cryptographic binding of manifest to receiver identity
& Private receiver policy is still not disclosed or negotiated \\
State authority spoofing & Attacker points evaluator to an unrecognized
or malicious state source & Signed \texttt{stateAuthorityPointer},
trust-registry validation, fail-closed on unrecognized authority &
Issuer-selected low-assurance state authorities create synchronization
risk \\
Stale cumulative state & Agent presents old state voucher or stale quota
evidence & Freshness and monotonicity invariant, \(\Delta t\) envelope,
signed audit record & Lag within accepted \(\Delta t\) is a governance
tradeoff \\
Request-context manipulation & Attacker alters local context fields used
for evaluation & Receiver-side enforcement, trusted request context,
fail-closed on missing fields & If the receiver's local context is
compromised, the model cannot correct it \\
Audit suppression & Enforcer allows action but omits signed audit record
& Mandatory signed audit record requirement & Enforcement engines must
be certified or monitored for conformance \\
\end{longtable}
}

The model therefore reduces ambiguity and replay risk, but it does not
eliminate the need for operational security around issuers, receivers,
registries, and state authorities. A compromised issuer key, malicious
receiver, corrupted local request context, or governance failure in a
trust registry remains outside the ability of the authorization
credential alone to repair. The purpose of the model is to make such
failures detectable, attributable, and fail-closed where possible.

\subsubsection{7.17 Privacy
Considerations}\label{privacy-considerations}

A portable authorization credential is not only a security artifact; it
is also a privacy-bearing artifact. A rich signed payload may reveal the
agent's identity, issuer relationship, delegated role, transaction
limits, permitted counterparties, workflow identifiers, trust anchors,
or state authority references. Across organizational boundaries, these
fields can expose more business context than the receiver strictly needs
in order to make the current authorization decision. Privacy therefore
has to be treated as part of the credential and profile design, not as
an afterthought.

The first privacy risk is \textbf{credential correlation}. If the same
credential identifier, subject identifier, proof-of-possession key,
workflow identifier, or distinctive constraint set is reused across
multiple receivers, colluding receivers may be able to link the agent's
activity across otherwise separate contexts. Audience binding helps, but
it is not always sufficient by itself. Profiles should prefer
receiver-scoped credentials, short-lived presentation artifacts,
audience-specific proof keys, and pairwise or scoped subject identifiers
where business accountability permits them. The goal is to let the
receiver verify the authority needed for this interaction without
turning the credential into a universal tracking handle.

As noted in the W3C Verifiable Credentials Data Model 2.0 privacy
considerations, stable identifiers and proof artifacts can enable
identifier-based and signature-based correlation across verifiers
{[}7{]}. VC-based authorization profiles should therefore prefer
receiver-scoped issuance, short lifetimes, minimized disclosure, and
audit records that retain credential digests or normalized evaluation
facts rather than storing the full credential by default.

The second risk is \textbf{over-disclosure}. The authorization payload
should carry only the facts required for the receiver's enforcement
decision. A receiver that needs to know that a settlement is below USD
5,000 does not necessarily need the delegator's broader budget, internal
approval hierarchy, or unrelated permissions. The pre-flight discovery
mechanism in Section 7.13 supports this by allowing the sender to
determine which profiles, fields, and trust anchors the receiver accepts
before presenting an authorization credential. That discovery step
should be used to select the narrowest sufficient credential or
attenuation, not the most information-rich credential available.

The third risk is \textbf{audit over-retention}. Signed audit records
are necessary for accountability, but they may also preserve sensitive
context long after the transaction has completed. Implementations should
distinguish between fields required for non-repudiation and dispute
resolution, and fields that are merely convenient for debugging.
Retention windows, access controls, redaction policies, and
jurisdictional requirements should be part of the applicable profile or
governance contract.

Selective-disclosure and zero-knowledge techniques may reduce
over-exposure in advanced profiles. For example, a credential holder may
be able to prove that a trusted issuer granted authority for a bounded
action, that a monetary value is within an authorized range, or that a
recipient belongs to an approved category without revealing unrelated
claims. The core model does not require a specific selective-disclosure
mechanism, because doing so would bind the authorization semantics to a
particular cryptographic construction. It does, however, require that
any such profile preserve deterministic evaluation: the receiver must
still obtain, or be able to verify, the typed facts needed for semantic
resolution, constraint evaluation, auditability, and fail-closed
behavior. If a privacy-preserving proof hides a field that the evaluator
must resolve in order to decide, the evaluator must deny rather than
guess.

{\def\LTcaptype{none} 
\begin{longtable}[]{@{}
  >{\raggedright\arraybackslash}p{(\linewidth - 4\tabcolsep) * \real{0.3333}}
  >{\raggedright\arraybackslash}p{(\linewidth - 4\tabcolsep) * \real{0.3333}}
  >{\raggedright\arraybackslash}p{(\linewidth - 4\tabcolsep) * \real{0.3333}}@{}}
\toprule\noalign{}
\begin{minipage}[b]{\linewidth}\raggedright
Privacy Concern
\end{minipage} & \begin{minipage}[b]{\linewidth}\raggedright
Example
\end{minipage} & \begin{minipage}[b]{\linewidth}\raggedright
Mitigation Direction
\end{minipage} \\
\midrule\noalign{}
\endhead
\bottomrule\noalign{}
\endlastfoot
Cross-receiver correlation & Same credential or subject identifier
reused at multiple receivers & Audience-specific credentials, pairwise
identifiers, short-lived presentations \\
Over-broad payload & Credential reveals unrelated roles, limits, or
internal approval structure & Data minimization, attenuation,
profile-specific required fields \\
Discovery leakage & Sender reveals sensitive capability information
during compatibility checking & Public receiver manifests, sender-side
local comparison, no private policy negotiation \\
Audit over-retention & Audit record stores more business context than
needed for later dispute resolution & Profile-defined retention, access
control, redaction, jurisdictional controls \\
Privacy-preserving proof mismatch & Selective-disclosure proof hides a
field required for evaluation & Typed disclosure commitments or
fail-closed denial \\
\end{longtable}
}

\paragraph{7.17.1 VC Profile
Considerations}\label{vc-profile-considerations}

VC-based profiles are a valid container option for this model,
especially where portable issuer verifiability and decentralized trust
resolution are useful across trust boundaries. Some privacy concerns
discussed in the VC ecosystem arise most strongly in human-centric
credential settings and are less acute in enterprise agent
authorization, where auditability, issuer accountability, and
non-repudiation are often intentional design goals rather than side
effects to be minimized at all costs.

Even so, VC profiles introduce real operational tradeoffs. Stable
identifiers and proof artifacts can still create correlation risk if
credentials are reused broadly across receivers. Status and revocation
mechanisms introduce additional lifecycle complexity. JSON-LD and Data
Integrity deployments may also impose higher verification overhead than
JWT/JWS-based profiles. These are profile-level tradeoffs, not
objections to the authorization model itself. Deployments that
prioritize lower operational overhead may prefer JWT/JWS profiles, while
deployments that need stronger cross-boundary portability or
decentralized trust resolution may prefer VC profiles.

\subsection{8. Discussion and Related
Work}\label{discussion-and-related-work}

\subsubsection{8.1 Related Work}\label{related-work}

The problem of agent authorization across trust boundaries has attracted
growing attention. Several recent proposals address fragments of the
challenge; none addresses the full surface described here.

\textbf{A-JWT} {[}10{]} extends JWT-based agent authorization with
intent binding and delegation-oriented claims. It is valuable for
improving traceability of who authorized a given call and under what
asserted context. The model proposed here differs in focus: rather than
centering primarily on invocation-bound delegation, it defines a
portable authorization payload with typed constraints, explicit
attenuation semantics, and shared cross-receiver evaluation behavior.

\textbf{Authenticated delegation for AI agents} {[}28{]} proposes an
OAuth/OpenID Connect-oriented framework in which users delegate
authority to AI agents through agent-specific credentials, delegation
tokens, scope restrictions, and audit metadata. That work is closely
aligned with the delegation and accountability problem space addressed
here, but it leaves the portable authorization semantics largely to the
credential or access-control profile. The model proposed in this paper
is complementary: it defines the common semantic layer that such
delegation credentials would need to carry and preserve across
containers, receivers, and trust boundaries.

\textbf{Recent agent identity protocol efforts} {[}9{]}, including those
built around Biscuit {[}12{]}- or Datalog-style embeddings, recognize
the need for offline attenuation, multi-hop delegation, and
provenance-aware evidence. Their strength lies in expressive delegation
and policy representation. The model proposed here makes a different
tradeoff: it constrains the core authorization algebra more tightly in
order to improve deterministic interoperability across independent
receivers, so that ecosystems need not align on a general-purpose
embedded logic in addition to a common artifact format.

{\def\LTcaptype{none} 
\begin{longtable}[]{@{}
  >{\raggedright\arraybackslash}p{(\linewidth - 4\tabcolsep) * \real{0.3333}}
  >{\raggedright\arraybackslash}p{(\linewidth - 4\tabcolsep) * \real{0.3333}}
  >{\raggedright\arraybackslash}p{(\linewidth - 4\tabcolsep) * \real{0.3333}}@{}}
\toprule\noalign{}
\begin{minipage}[b]{\linewidth}\raggedright
\textbf{Dimension}
\end{minipage} & \begin{minipage}[b]{\linewidth}\raggedright
\textbf{Logic-Embedded Approaches}
\end{minipage} & \begin{minipage}[b]{\linewidth}\raggedright
\textbf{Proposed Model}
\end{minipage} \\
\midrule\noalign{}
\endhead
\bottomrule\noalign{}
\endlastfoot
Policy expressiveness & High & Deliberately constrained \\
Interoperability burden & Requires alignment on evaluation logic &
Requires alignment on constrained shared semantics \\
Best fit & Rich local or ecosystem-specific policy environments &
Portable cross-receiver authorization with shared evaluation
semantics \\
\end{longtable}
}

\textbf{UCAN} {[}11{]} provides a strong capability-based model for
delegation, attenuation, and cryptographic chaining. Its main focus,
however, is capability transfer rather than a standardized constraint
algebra for enterprise-style authorization conditions such as monetary
bounds, temporal windows, recipient restrictions, or cumulative exposure
controls. The model proposed here is intended to address that more
constrained but more portable cross-receiver authorization problem.

\textbf{zCAP-LD} {[}23{]}, {[}24{]} is especially relevant because it
makes the separation between an authorization envelope and a caveat
language explicit. That architectural move aligns with this paper's
separation of container from authorization meaning. However, zCAP-LD
remains an incubating W3C Credentials Community Group work item as of
April 2026 rather than a finalized global standard, and its center of
gravity is still capability delegation and invocation. This paper
therefore treats zCAP-LD as informative prior art and as a possible
future container/profile option, not as a normative dependency.

\textbf{XACML} {[}13{]}, \textbf{OPA} {[}14{]}, and \textbf{Cedar}
{[}15{]} are highly relevant because they demonstrate that fine-grained
policy evaluation is operationally feasible. They are primarily oriented
around receiver-authored policy evaluation, however, rather than around
a standardized, issuer-authored portable authorization payload that
independent conformant receivers can use to reach consistent
authorization decisions without bespoke bilateral policy logic. The gap
addressed here is therefore not whether these systems can perform
fine-grained evaluation, but whether they define a portable policy
payload with shared semantics across independent receivers.

\textbf{ODRL} {[}22{]} is relevant as a standardized policy model and
vocabulary. The model proposed here does not adopt ODRL's full policy
algebra, rule taxonomy, or logical combinators as normative
requirements. Instead, selected ODRL-style constraint terms could be
used in an optional serialization or compatibility profile where they
preserve the conjunctive, total, and fail-closed evaluation semantics
defined in this paper. ODRL is therefore best understood here as a
possible vocabulary-alignment path, not as a replacement for the core
authorization model.

\textbf{OpenID AuthZEN} {[}21{]} is an important adjacent effort because
it standardizes the interface between a Policy Enforcement Point and a
Policy Decision Point. Its focus is the authorization API: how a PEP
asks for an access decision, how the PDP returns that decision, and how
PDP metadata and capabilities can be discovered. This paper is
complementary rather than competing. AuthZEN can standardize the
decision-call boundary inside or across systems, while the model
proposed here standardizes what bounded authority an agent carries, how
that authority is expressed, and how independent receivers evaluate it
deterministically across trust boundaries.

\textbf{SPIFFE and workload identity systems} {[}18{]} provide strong
presenter identity within and across trust domains, but they do not
themselves define portable authorization semantics. They are therefore
complementary: presenter authentication can be handled by SPIFFE-style
mechanisms, while authorization is carried in the portable authorization
credential defined here.

\subsubsection{8.2 Liability and
Accountability}\label{liability-and-accountability}

The authorization semantic model does not, and cannot, resolve the
underlying legal question of liability when an autonomous agent causes
financial or operational harm. Liability remains a matter of contract
law, terms and conditions, and established regulatory frameworks. What
the model does provide is the evidentiary infrastructure without which
liability determination in multi-agent systems becomes speculative or
impossible.

In traditional enterprise workflows, a human signature or manual
approval provides a clear point of accountability. In agentic systems,
the responsibility chain must instead be reconstructed from digital
artifacts: who empowered the agent, what constraints were in force, what
request was made, what evaluation occurred, and what outcome was
produced. The signed credential, the evaluation trace, and the signed
audit record together provide this reconstruction. However, for that
reconstruction to be legally and operationally meaningful, the model
must also account for the security context in which the agent was
authenticated and operated.

\paragraph{8.2.1 Capturing Security Context for Reasonable
Care}\label{capturing-security-context-for-reasonable-care}

To support dispute resolution and post-incident analysis, the signed
audit trail should capture not only the agent's identity and
authorization state, but also relevant technical metadata about how the
agent was authenticated and protected at the time of action. This
creates a factual basis for determining whether the principal i.e., the
user, enterprise, or delegated authority behind the agent exercised
reasonable care in securing that agent.

A first distinction concerns \textbf{hardware versus software binding}.
An audit record should be able to indicate whether the agent's signing
keys or credentials were protected by stronger controls such as a
hardware security module, trusted execution environment, or another
rooted hardware boundary, or whether they relied on software-based
storage alone. This distinction matters when an incident later raises
the question of whether the failure arose from flawed business logic,
weak operational controls, or credential compromise.

A second concern is \textbf{credential integrity and compromise
posture}. In the event of unauthorized spend, fraudulent approval, or
improper evidence disclosure, technical metadata about the credential's
protection state allows investigators, insurers, courts, or regulators
to distinguish between a systemic decision failure and a security
breach. An agent that acted within its signed policy but under flawed
human-defined logic presents a different accountability picture from an
agent whose credentials were exfiltrated because its keys were poorly
protected.

A third concern is \textbf{attestation of environment}. Where available,
the audit record should include signed attestations about the agent's
execution environment, such as whether the agent was running in an
expected workload, trusted runtime, or verified platform state at the
time the action occurred. This extends the evidentiary function of the
audit trail beyond the decision itself and into the surrounding trust
state of the system that produced it.

By formalizing the capture of this security context, the model does not
attempt to automate liability allocation. Rather, it strengthens the
evidentiary basis on which liability, negligence, breach, or compliance
failure may later be assessed. In a cross-boundary agent ecosystem, this
level of technical transparency is likely to be necessary if
organizations are to enter into the contractual and governance
arrangements that meaningful autonomous collaboration will require.

\paragraph{8.2.2 Protocol Conformance and Liability Partition Under
Infrastructure
Lag}\label{protocol-conformance-and-liability-partition-under-infrastructure-lag}

The model does not determine ultimate legal liability, but it does
define a protocol-level notion of reasonable care. In this context,
reasonable care is not a guarantee of perfect outcomes. It is
conformance to the active enforcement contract: the applicable
governance manifest, trust-registry requirements, permitted state
authority model, profile-defined freshness envelope \(\Delta t\), and
mandatory audit obligations.

This matters when the protocol is followed correctly but the business
outcome is still harmful because of infrastructure lag. If the issuer
binds a permitted \texttt{stateAuthorityPointer} into the credential, it
is appointing that state source as authoritative for cumulative
governance under the applicable profile. If an enforcer verifies
freshness within the active \(\Delta t\), a harmful outcome caused by
synchronization lag within that envelope should not by itself imply
enforcer negligence. In protocol terms, the enforcer acted conformantly
against the authoritative source and freshness assumptions accepted by
the governance model.

Responsibility shifts when the enforcer bypasses those safety gates. If
the audit trail shows that the enforcer accepted an unpermitted state
authority, ignored state older than the active \(\Delta t\), or failed
to produce the required signed audit record, the failure is no longer
attributable merely to infrastructure lag. It becomes a failure of
protocol conformance.

{\def\LTcaptype{none} 
\begin{longtable}[]{@{}
  >{\raggedright\arraybackslash}p{(\linewidth - 4\tabcolsep) * \real{0.3333}}
  >{\raggedright\arraybackslash}p{(\linewidth - 4\tabcolsep) * \real{0.3333}}
  >{\raggedright\arraybackslash}p{(\linewidth - 4\tabcolsep) * \real{0.3333}}@{}}
\toprule\noalign{}
\begin{minipage}[b]{\linewidth}\raggedright
Failure Scenario
\end{minipage} & \begin{minipage}[b]{\linewidth}\raggedright
Presumptive Responsibility Under the Enforcement Contract
\end{minipage} & \begin{minipage}[b]{\linewidth}\raggedright
Primary Evidence Source
\end{minipage} \\
\midrule\noalign{}
\endhead
\bottomrule\noalign{}
\endlastfoot
Credential exfiltration or key compromise & Issuer / principal-side
security failure & Technical protection metadata, revocation records \\
Logic or policy flaw in signed authority & Issuer / delegating authority
& Signed authorization payload, evaluation trace \\
Synchronization lag within permitted \(\Delta t\) & Issuer-selected
governance model / appointed state authority & Voucher timestamp,
profile-defined \(\Delta t\), audit trail \\
Enforcer bypasses required state check & Enforcer non-conformance &
Audit trail \\
Enforcer accepts stale state outside permitted \(\Delta t\) & Enforcer
non-conformance & Audit trail, voucher timestamp, profile-defined
\(\Delta t\) \\
Enforcer accepts unpermitted state authority & Enforcer non-conformance
& Trust-registry verification record, audit trail \\
Missing signed audit record & Enforcer non-conformance & Absence of
required audit artifact \\
\end{longtable}
}

\subsubsection{8.3 Temporal Model
Considerations}\label{temporal-model-considerations}

The two use cases examined in this paper expose a tension in temporal
semantics. Insurance claims processing is predominantly point-in-time: a
request arrives, the evaluator checks whether the current moment falls
within the credential's temporal window, and a decision is made.

Supply-chain integrity workflows unfold over longer periods, and the
relevant question may sometimes be not only whether a request is
currently valid, but whether a given attestation was valid at the time
it was created or transferred.

The model proposed here directly addresses the first of these temporal
dimensions. It supports point-in-time authorization for requesting,
relaying, and presenting evidence, while preserving provenance,
delegation integrity, and auditability across those interactions.
Longer-horizon adjudication of historical evidentiary validity
introduces an additional temporal layer that may be better served by
profile extensions such as shared evidence logs, verifiable historical
status records, or other cross-domain provenance mechanisms, without
changing the core evaluation model.

\subsubsection{8.4 Semantic Mapping Risk}\label{semantic-mapping-risk}

Governed semantic mapping introduces a category of risk that does not
arise when both parties use identical, pre-agreed field names. If a
credential constrains one semantic identifier but the receiver maps it
incorrectly to a local field, the evaluator may execute correctly
against the wrong resolved input. This is the central tradeoff of
semantic portability across heterogeneous systems.

The alternative would be to require all issuers and receivers to adopt
identical field names and internal schemas. This would sharply limit
portability and reduce adoption to tightly pre-negotiated bilateral
relationships. The mitigation is therefore not to abandon semantic
mapping, but to govern it strictly through versioned profiles, validated
aliases, type-safe coercion, and fail-closed enforcement, while allowing
domain vocabularies to converge over time.

\subsubsection{8.5 Stateful Governance and Enforcement
Contracts}\label{stateful-governance-and-enforcement-contracts}

The model focuses primarily on stateless evaluation in order to
prioritize high-performance, deterministic outcomes without requiring a
synchronous dependency on a global state store. This design choice is
useful for portability and predictable enforcement. For enterprise-grade
deployments, particularly in financial services and high-velocity
coordination settings, however, it also leaves a cumulative gap. A
purely per-action ceiling constrains each individual action but does not
by itself bound aggregate exposure across a sequence of otherwise valid
actions, leaving organizations exposed to substantial financial and
operational risk.

\paragraph{8.5.1 The Enforcement Contract: Mutual Protection and
Reasonable
Care}\label{the-enforcement-contract-mutual-protection-and-reasonable-care}

To address this, stateful constraints can be understood as part of an
enforcement contract between the issuer and the enforcer. This framing
does not attempt to redefine legal liability. Rather, it captures the
idea that both parties have a due-diligence interest in ensuring that
authority remains bounded over time. The issuer uses cumulative or
rate-aware constraints to communicate a meaningful stop-loss boundary.
The enforcer, by respecting those limits, demonstrates that it acted
with reasonable care rather than treating a sequence of individually
valid requests as unlimited aggregate authority.

In this sense, stateful governance serves as mutual protection. For the
issuer, a cumulative limit helps prevent treasury depletion, rolling
budget overrun, or runaway exposure caused by individually compliant but
collectively excessive actions. For the enforcer, respecting such limits
provides a defensible basis for transaction clearance, payout control,
infrastructure protection, and post-incident review.

{\def\LTcaptype{none} 
\begin{longtable}[]{@{}
  >{\raggedright\arraybackslash}p{(\linewidth - 4\tabcolsep) * \real{0.3333}}
  >{\raggedright\arraybackslash}p{(\linewidth - 4\tabcolsep) * \real{0.3333}}
  >{\raggedright\arraybackslash}p{(\linewidth - 4\tabcolsep) * \real{0.3333}}@{}}
\toprule\noalign{}
\begin{minipage}[b]{\linewidth}\raggedright
\textbf{Constraint Type}
\end{minipage} & \begin{minipage}[b]{\linewidth}\raggedright
\textbf{Issuer Motivation (Risk Control)}
\end{minipage} & \begin{minipage}[b]{\linewidth}\raggedright
\textbf{Enforcer Motivation (Due Diligence)}
\end{minipage} \\
\midrule\noalign{}
\endhead
\bottomrule\noalign{}
\endlastfoot
Numeric Limit & Prevents total treasury depletion & Ensures transaction
clearance remains within authorized mandate \\
Temporal Window & Ensures operational and policy alignment & Confirms
regulatory and audit compliance \\
Enumerated List & Restricts categories, recipients, or jurisdictions &
Prevents out-of-scope handling or disclosure \\
String Pattern & Bounds resource namespace and evidence scope & Prevents
unauthorized traversal across resource domains \\
Velocity / Rate & Guards against algorithmic runaway behavior & Protects
infrastructure from unintended denial-of-service patterns \\
\end{longtable}
}

\paragraph{\texorpdfstring{8.5.2 Proposed Extension:
\texttt{Cumulative\-Limit\-Constraint}}{8.5.2 Proposed Extension: CumulativeLimitConstraint}}\label{proposed-extension-cumulativelimitconstraint}

Unlike the stateless constraints described in Section 7.3, the
\texttt{Cumulative\-Limit\-Constraint} is designed to govern aggregate risk
across an otherwise unbounded sequence of autonomous actions. This
extension is best understood as an advanced profile layered on top of
the stateless core, not as a requirement of the normative model itself.
This constraint type is not self-contained. Instead, it binds the
authorization grant to a specific external source of truth through a
mandatory \texttt{stateAuthorityPointer}.

The \texttt{stateAuthorityPointer} is a cryptographically signed URI or
equivalent identifier embedded within the credential that points to the
authoritative state store. Depending on the deployment profile, this may
reference a balance registry, a distributed ledger entry, a governed
high-performance counter, or another trusted state authority.

Evaluation therefore proceeds in two stages. First, the enforcement
engine validates the individual request against any applicable
per-action ceiling. Second, it queries the referenced
\texttt{stateAuthorityPointer} to determine whether the cumulative value
after the requested action remains within the authorized aggregate
budget for the relevant period.

The integrity requirements remain fail-closed. If the referenced state
authority is unreachable, stale beyond the allowed threshold, or
otherwise unverifiable, the evaluator must deny the request. The
evaluator must also validate that the referenced
\texttt{stateAuthorityPointer} is permitted by the relevant
trust-registry entry described in Section 7.6.1.1. This ensures that
autonomous action does not proceed under conditions of unresolved
financial or operational uncertainty or against an unrecognized state
authority.

This structure makes an important separation explicit: the
\textbf{what}, the cumulative limit itself, and the \textbf{where}, the
authoritative state pointer, are signed together inside the credential.
The question of \textbf{who is allowed to host that state authority} is
governed externally through the trust-registry framework described in
Section 7.6.1.1.

\begin{center}
\includegraphics[width=0.9\textwidth]{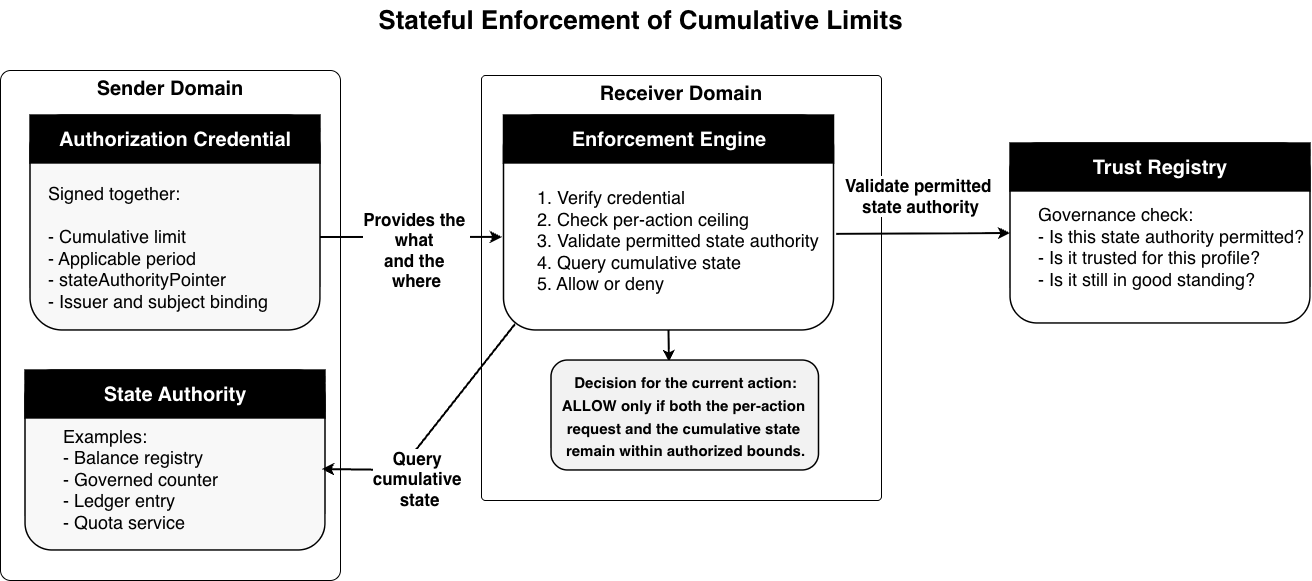}
\end{center}

\emph{Figure 9: Stateful enforcement of cumulative limits through an
external state authority. The credential signs both the cumulative limit
and the pointer to the authoritative state source; the enforcement
engine evaluates both the current action and the cumulative state, while
the trust registry governs which state authorities are acceptable.}

\paragraph{8.5.3 Verifiable State Proofs}\label{verifiable-state-proofs}

To maintain portability in decentralized or multi-party environments
where a persistent connection to a \texttt{stateAuthorityPointer} is not
guaranteed, the model can support verifiable state proofs as an advanced
profile.

Under this pattern, the agent carries a cryptographically signed state
voucher that serves as an attestation of its current spent balance or
remaining quota. Upon each successful request, the enforcer updates the
voucher and signs a new balance proof. A subsequent enforcer verifies
the chain of state updates before accepting the voucher as the current
state. This makes it more difficult for an agent to replay an older,
more permissive state artifact in an attempt to bypass the
\texttt{Cumulative\-Limit\-Constraint}.

Verifiable state proofs therefore offer a decentralized analogue to a
centralized balance registry, while preserving cryptographic continuity
across multiple enforcers. This mechanism is best viewed as an advanced
profile rather than a mandatory core requirement. It introduces
additional considerations around replay resistance, ordering, and proof
freshness, but it provides a viable path for cumulative governance in
environments where centralized state coordination is impractical.

\begin{center}
\includegraphics[width=0.9\textwidth]{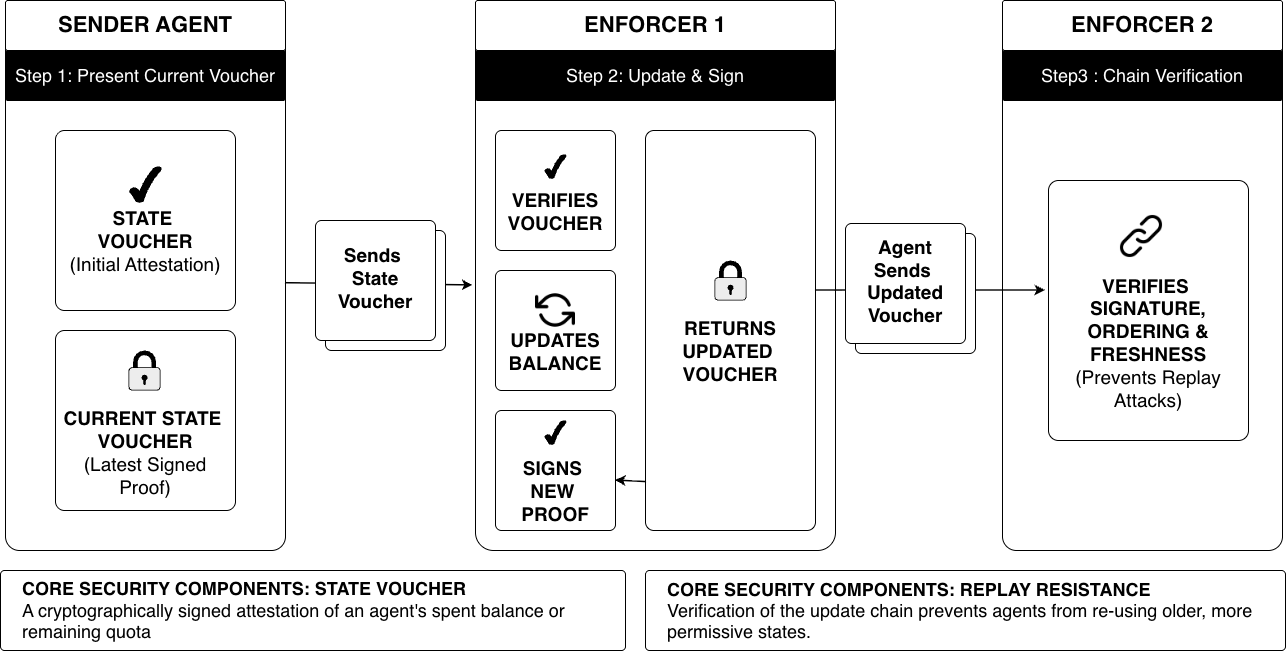}
\end{center}

\emph{Figure 10: Verifiable state proofs as an advanced profile for
cumulative governance. An agent carries the latest signed state voucher;
each enforcer verifies the current voucher, updates the state, signs a
new proof, and later enforcers verify signature continuity, ordering,
and freshness to resist replay.}

\paragraph{8.5.3.1 Freshness and Monotonicity
Invariant}\label{freshness-and-monotonicity-invariant}

To resist replay in asynchronous environments, verifiable state proofs
must preserve both freshness and monotonicity. A conformant
\texttt{StateVoucher} therefore carries three elements: a monotonic
sequence value, a signed timestamp, and a signature chain linking the
current voucher to previously accepted state transitions.

These three elements serve distinct but complementary purposes. The
monotonic sequence value protects ordering and prevents rollback or
replay of older vouchers. The timestamp, interpreted at the granularity
defined by the applicable profile, allows the evaluator to test
freshness within a bounded drift window. UTC normalization, or an
equivalently unambiguous time representation, prevents ambiguity arising
from local time-zone conventions, daylight saving time transitions, or
inconsistent clock formats across domains.

An enforcer's evaluation of the \texttt{Cumulative\-Limit\-Constraint} in
Section 8.5.2 is valid only if three conditions hold. First,
\textbf{sequential continuity}: the presented sequence value must extend
the accepted voucher chain monotonically and must not duplicate or roll
back a previously accepted state. Any reused, duplicated, or lower
sequence value must be denied. Second, \textbf{temporal proximity}: the
voucher timestamp must fall within the maximum drift envelope
\(\Delta t\) defined by the active industry profile. Its purpose is
bounded freshness validation, not mere calendrical description. The
applicable profile should also define the required timestamp
granularity, such as seconds or milliseconds, based on deployment
sensitivity. If the voucher age exceeds \(\Delta t\), the evaluator must
fail closed and return a typed denial reason indicating stale state.
Third, \textbf{authority continuity}: the voucher must carry a valid
signature from a prior enforcer or state authority recognized by the
relevant trust registry and permitted for that profile. This prevents
acceptance of vouchers signed by unauthorized intermediaries.

Together, these conditions allow verifiable state proofs to preserve
replay resistance and continuity of state without requiring a
permanently online shared state authority. Recovery or reconciliation
after failure may be handled through external governance mechanisms, but
the authorization decision for the current interaction remains deny.

\paragraph{8.5.4 Epoch-Based Distributed
Quotas}\label{epoch-based-distributed-quotas}

A more scalable pattern for distributed environments is epoch-based
quota allocation. Rather than querying or updating a global state
authority on every request, the issuer pre-allocates bounded spending
windows (for example, daily or hourly sub-limits) to individual
enforcers or execution clusters. Each enforcer then operates without
synchronous cross-cluster coordination on each request until the current
epoch ends, at which point quotas are refreshed.

This preserves the high-throughput, low-latency properties of the core
model while reducing coordination overhead. It does introduce bounded
temporary over-allocation risk if multiple enforcers are given quota
slices simultaneously, but that risk is explicit, modeled, and
operationally governable. Like verifiable state proofs, epoch-based
quotas are best treated as a deployment profile layered on top of the
stateless core rather than as a mandatory requirement of the standard.

\paragraph{8.5.5 Enforcement Tiers and Availability
Tradeoffs}\label{enforcement-tiers-and-availability-tradeoffs}

Enterprises may configure enforcement engines to apply different tiers
of stateful validation based on the risk profile of the transaction.

A \textbf{stateless tier} is appropriate for lower-risk actions and
relies only on per-action evaluation. It maximizes portability and
availability but does not address cumulative exposure.

An \textbf{epoch-bound tier} applies local quota enforcement with
periodic synchronization. It provides stronger macro-governance while
retaining acceptable performance for many operational settings.

A \textbf{synchronous tier} requires real-time verification against the
authoritative state source for every high-risk action. It offers the
strongest control but at the cost of availability and latency.

By standardizing how the enforcement engine interacts with the relevant
state source, the model preserves its portable core while supporting the
stronger governance required for high-risk autonomous coordination.

\subsubsection{8.6 Empirical Validation and Reference
Implementation}\label{empirical-validation-and-reference-implementation}

This paper presents the authorization semantic model as a formal
specification with pseudocode evaluation procedures. Empirical
validation remains future work. In particular, three areas require
implementation experience before standardization.

\textbf{Reference implementation.} A minimal open-source enforcement
engine implementing the canonical processing pipeline in Section 7.11,
semantic resolution in Section 7.12.5, and delegation-chain verification
would allow independent parties to test interoperability claims against
concrete artifacts rather than prose descriptions alone.

\textbf{Performance characterization.} Constraint evaluation over the
four core types is designed to be local and \texttt{O(n)} in the number
of constraints, but real-world overhead depends on container parsing,
signature verification, governance-manifest fetching, and, for stateful
tiers, state-authority round trips. Benchmarking these stages across
JWT/JWS and VC containers under representative workloads would quantify
the latency budget and inform deployment guidance.

\textbf{Conformance test vectors.} The worked trace in Section 7.11.1
demonstrates one allow path and one deny path. A conformance suite
should cover every denial reason enumerated in the pseudocode, including
\texttt{issuer\_untrusted}, \texttt{audience\_mismatch},
\texttt{delegation\_widened}, \texttt{constraint\_unknown}, and
\mbox{\texttt{context\_field\_missing}}, to verify that independent
implementations converge on identical decisions for the same inputs.

These artifacts are natural companion deliverables alongside the
standardization path described in Section 9.

\subsection{9. Toward a Standard}\label{toward-a-standard}

The authorization semantic model described in Section 7 is intended to
serve as the foundation for a formal standard. A practical
standardization path should distinguish among a normative core, profile
bindings, governed vocabularies, mapping profiles, and optional advanced
governance profiles. This separation keeps the core portable and
testable while allowing container-specific, industry-specific, and
stateful extensions to evolve without fragmenting the model.

No single standards body needs to own every layer. A practical path
would likely split the work across venues according to their strengths:
IETF-style work for compact signed artifacts, JWT/JWS bindings,
proof-of-possession, and well-known discovery mechanics; the OpenID
Foundation, particularly adjacent to AuthZEN-style authorization
interoperability, for authorization API and PDP/PEP integration
profiles; the W3C Verifiable Credentials ecosystem for VC packaging and
credential-status alignment; and industry consortia or regulated-sector
governance bodies for vertical vocabularies such as insurance, financial
services, healthcare, or aerospace supply-chain profiles. The important
point is not institutional exclusivity, but artifact separation: each
layer should be independently versioned, testable, and adoptable.

\subsubsection{9.1 Normative Core}\label{normative-core}

The normative core should define the authorization payload structure,
mandatory payload components, the core constraint types and their
evaluation and attenuation semantics, the conjunctive total fail-closed
evaluation model, canonical typing and mapping rules, the
most-restrictive-wins merge rule, delegation-chain verification, and
signed audit-record requirements.

Two conformant implementations, given the same signed authorization,
request context, applicable profile and vocabulary versions, and
canonical typing and mapping rules, should produce the same
authorization decision. This implies the need for a conformance test
suite containing machine-readable test vectors that cover successful
authorization, failure cases for each constraint type, fail-closed
behavior on unknown constraints, missing context fields, local-policy
restriction, delegation attenuation, revocation, expiration, issuer
trust verification, proof of possession, subject binding, and audience
binding.

\subsubsection{9.2 Profile Bindings}\label{profile-bindings}

Profile bindings should specify how the authorization payload maps into
a particular credential container and associated trust model.

A JWT profile {[}1{]} binding would map agent identity, issuer identity,
permissions, structured constraints, audience binding, and
proof-of-possession requirements into claims and associated validation
rules.

A W3C Verifiable Credential profile {[}7{]} binding would map the same
semantic components into VC structures, proof methods, and status
mechanisms.

An OAuth Rich Authorization Requests profile {[}3{]} could similarly
carry the authorization payload through the
\mbox{\texttt{authorization\_details}} structure. Work on a Cedar profile for
OAuth Rich Authorization Requests {[}27{]} illustrates this direction by
defining a way to distribute Cedar policy sets through
\mbox{\texttt{authorization\_details}}. In the architecture proposed here, such
work is best understood as a concrete profile binding: OAuth RAR
provides the container and exchange surface, while Cedar provides one
possible policy-language serialization and enforcement substrate. The
portable authorization semantics, Minimum Viable Vocabulary, attenuation
rules, governed mapping behavior, audit requirements, and conformance
expectations remain part of the common semantic layer that a profile
binding must preserve.

The critical requirement is that different bindings preserve the same
authorization meaning. Profile choice may change serialization, trust
bootstrapping, and revocation method, but it must not alter the
authorization decision when the same signed authority, request context,
and applicable vocabulary semantics are presented to a conformant
evaluator.

\subsubsection{9.3 Governed Vocabularies and Mapping
Profiles}\label{governed-vocabularies-and-mapping-profiles}

Governed vocabularies should define permission namespaces, resource and
action naming conventions, required context fields, field types, and
approved aliases for a specific industry or use case. The Minimum Viable
Vocabulary provides the shared semantic substrate for portable
authorization, while domain profiles extend that substrate under
explicit governance and versioning.

Mapping profiles are equally important. They define how local request
structures resolve to shared semantic identifiers and which type-safe
coercions are permitted. These mappings are not ad hoc runtime
inference; they are governed artifacts that enable semantic portability
across heterogeneous systems. Signed discovery metadata, such as the
receiver governance manifest described in Section 7.13, may publish
accepted vocabularies, profile versions, trust anchors, and required
context fields in advance, in order to reduce integration brittleness
while preserving authoritative receiver-side enforcement.

Stateful governance extensions such as cumulative limits, epoch-bound
quotas, and verifiable state proofs should be treated as optional
advanced profiles layered on top of the normative core rather than as
mandatory features for every conformant implementation.

\subsubsection{9.4 Conformance and
Interoperability}\label{conformance-and-interoperability}

A conformance program should include at least four layers.

\textbf{Level 1 Evaluation Conformance:} The implementation correctly
evaluates the core constraint types, applies conjunctive total
fail-closed logic, and produces the expected decisions for the normative
test vectors.

\textbf{Level 2 Semantic Conformance:} The implementation correctly
recognizes the Minimum Viable Vocabulary, applies governed mapping
profiles, enforces type-safe resolution, and fails closed on ambiguous,
stale, conflicting, or invalid semantic mappings.

\textbf{Level 3 Profile Conformance:} The implementation correctly
serializes and deserializes authorization payloads in at least one
supported profile and preserves semantic equivalence across encodings.

\textbf{Level 4 Delegation Conformance:} The implementation correctly
verifies delegation chains, enforces attenuation for all supported
constraint types, rejects widened delegations, and handles revocation
propagation.

\subsubsection{9.5 Concrete Work-Item
Structure}\label{concrete-work-item-structure}

A standards effort could be decomposed into a small number of concrete
work items rather than attempted as one large omnibus specification.

{\def\LTcaptype{none} 
\begin{longtable}[]{@{}
  >{\raggedright\arraybackslash}p{(\linewidth - 4\tabcolsep) * \real{0.3333}}
  >{\raggedright\arraybackslash}p{(\linewidth - 4\tabcolsep) * \real{0.3333}}
  >{\raggedright\arraybackslash}p{(\linewidth - 4\tabcolsep) * \real{0.3333}}@{}}
\toprule\noalign{}
\begin{minipage}[b]{\linewidth}\raggedright
Work Item
\end{minipage} & \begin{minipage}[b]{\linewidth}\raggedright
Purpose
\end{minipage} & \begin{minipage}[b]{\linewidth}\raggedright
Likely Home
\end{minipage} \\
\midrule\noalign{}
\endhead
\bottomrule\noalign{}
\endlastfoot
Core Authorization Semantics & Defines the payload model, MVV,
constraint algebra, evaluation semantics, attenuation rules, audit
record, and conformance vectors & IETF-style standards track or
comparable cross-vendor venue \\
JWT/JWS Profile & Defines compact enterprise-friendly serialization,
claims, audience binding, proof-of-possession, revocation hooks, and
validation rules & IETF or OpenID Foundation \\
Verifiable Credential Profile & Defines VC representation, proof method
expectations, status integration, and semantic equivalence with the core
payload & W3C Verifiable Credentials ecosystem \\
Governance Manifest Profile & Defines
\texttt{/.well-known/agent-governance}, signed manifest structure, cache
semantics, profile references, and compatibility metadata & IETF or
OpenID Foundation \\
Vocabulary and Mapping Profile Registry & Defines MVV registry rules,
profile identifiers, versioning metadata, alias rules, and
machine-readable mapping profile format & Shared registry with
industry-governed extensions \\
Insurance Profile & Defines claim, settlement, recipient, policy, and
fraud-review vocabularies for the insurance use case & Insurance
industry consortium or profile working group \\
Supply-Chain Evidence Profile & Defines part, shipment, attestation,
provenance, evidence disclosure, and recipient-role vocabularies &
Aerospace, defense, or supply-chain governance body \\
Stateful Governance Profile & Defines cumulative limits, state authority
pointers, epoch-bound quotas, state vouchers, and freshness invariants &
Advanced profile after the stateless core is stable \\
Conformance Test Suite & Provides executable test vectors for
evaluation, semantic resolution, profile bindings, delegation,
revocation, and failure cases & Cross-venue implementation project \\
\end{longtable}
}

This structure allows early adopters to implement a narrow, useful core
without waiting for every industry vocabulary or advanced stateful
mechanism to mature.

\subsubsection{9.6 Phased Adoption Path}\label{phased-adoption-path}

An adoption path should begin with the stateless core, because it offers
the highest interoperability value with the least operational
dependency.

\textbf{Phase 1: Core and One Container.} Define the core payload, MVV,
four constraint types, evaluation semantics, audit record, and a JWT/JWS
profile. Publish basic conformance vectors for allow, deny, expiration,
audience mismatch, unknown constraint, missing context, and local-policy
restriction.

\textbf{Phase 2: Semantic Portability.} Standardize the governance
manifest, mapping profile format, profile identifiers, and sender-side
pre-flight compatibility checks. Add conformance tests for semantic
resolution, stale manifests, alias conflicts, type mismatches, and
unsupported profile versions.

\textbf{Phase 3: Second Container and Early Industry Profiles.} Add the
VC profile and one or two concrete vertical profiles, such as insurance
settlement and supply-chain evidence. At this stage, interoperability
testing should show that the same authorization meaning survives across
JWT/JWS and VC containers.

\textbf{Phase 4: Delegation and Multi-Agent Workflow.} Expand
conformance testing for attenuation, delegation-chain verification,
revocation propagation, and multi-principal workflow composition. This
phase should also define clearer audit requirements for collaborative
agent workflows.

\textbf{Phase 5: Stateful Governance.} Introduce advanced profiles for
cumulative limits, epoch-bound quotas, state authority pointers, and
verifiable state proofs. This phase should remain optional until the
stateless and semantic layers have achieved implementation experience.

Taken together, this standardization path preserves a narrow, testable,
and portable core while still allowing richer industry semantics,
discovery contracts, and stateful governance mechanisms to evolve as
profiles. The goal is not to force all agent ecosystems into a single
trust architecture or runtime stack, but to define enough shared
authorization semantics that autonomous agents can carry bounded
authority across organizational boundaries in a way that remains
interoperable, auditable, and governable.

\subsubsection{9.7 Outstanding Design Considerations and Future
Work}\label{outstanding-design-considerations-and-future-work}

Several important areas warrant follow-on implementation work and
profile development. They do not weaken the normative core proposed
here, but they do shape how richer deployments may mature.

\begin{itemize}
\tightlist
\item
  \textbf{Revocation at scale.} The paper defines revocation
  semantically and identifies profile paths, including Bitstring Status
  List for VC profiles and TTL, introspection, or issuer-managed
  revocation for JWT/JWS profiles. Deeper implementation work remains
  for higher-complexity cases such as multi-hop delegated revocation,
  offline operation, privacy-preserving status checks, and revocation
  performance under high-volume cross-domain workloads.
\item
  \textbf{Event-driven security signaling.} Eventing frameworks such as
  OpenID Shared Signals Framework (SSF) {[}25{]} may complement this
  model by distributing revocation, compromise, or risk signals in near
  real time. Such mechanisms are relevant as operational overlays, but
  they do not replace signed portable authorization credentials or
  deterministic receiver-side evaluation semantics.
\item
  \textbf{Adjacent agent infrastructure ecosystems.} Emerging
  agent-infrastructure efforts such as AGNTCY {[}26{]} merit further
  evaluation as adjacent work, particularly where identity, discovery,
  messaging, and observability are being standardized together. They
  address a broader stack than the authorization model isolated in this
  paper, but practical interoperability between those layers remains
  important.
\item
  \textbf{Stateful governance non-functional requirements.} Section 8.5
  introduces advanced profiles for cumulative limits, epoch-bound
  quotas, and verifiable state proofs. Their latency, freshness, replay
  resistance, coordination cost, and failure-mode implications require
  reference implementation and deployment experience before they can be
  standardized confidently beyond the current architectural treatment.
\end{itemize}

\subsection{About the Author}\label{about-the-author}

Partha Madhira is a technology executive and architect focused on
enterprise AI systems, agentic architectures, digital identity, and
authorization models for autonomous systems. His work spans applied AI,
cloud architecture, governance, secure enterprise integration, and
technical community engagement.

\subsection{References}\label{references}

\begingroup\sloppy
\subsubsection{Normative References}\label{normative-references}

{[}1{]} M. Jones, J. Bradley, and N. Sakimura, ``JSON Web Token (JWT),''
RFC 7519, IETF, May 2015. Available:
https://www.rfc-editor.org/rfc/rfc7519

{[}2{]} D. Hardt, ``The OAuth 2.0 Authorization Framework,'' RFC 6749,
IETF, October 2012. Available: https://www.rfc-editor.org/rfc/rfc6749

{[}3{]} T. Lodderstedt, D. Fett, and J. Schaar, ``OAuth 2.0 Rich
Authorization Requests,'' RFC 9396, IETF, May 2023. Available:
https://www.rfc-editor.org/rfc/rfc9396

{[}4{]} M. Jones, A. Nadalin, B. Campbell, J. Bradley, and C. Mortimore,
``OAuth 2.0 Token Exchange,'' RFC 8693, IETF, January 2020. Available:
https://www.rfc-editor.org/rfc/rfc8693

{[}5{]} D. Fett, B. Campbell, J. Bradley, T. Lodderstedt, M. Jones, and
D. Waite, ``OAuth 2.0 Demonstrating Proof-of-Possession at the
Application Layer (DPoP),'' RFC 9449, IETF, September 2023. Available:
https://www.rfc-editor.org/rfc/rfc9449

{[}6{]} M. Jones, N. Sakimura, and J. Bradley, ``JSON Web Key (JWK),''
RFC 7517, IETF, May 2015. Available:
https://www.rfc-editor.org/rfc/rfc7517

{[}7{]} World Wide Web Consortium (W3C), ``Verifiable Credentials Data
Model v2.0,'' W3C Recommendation, 2024. Available:
https://www.w3.org/TR/vc-data-model-2.0/

{[}8{]} M. Nottingham, ``Well-Known Uniform Resource Identifiers
(URIs),'' RFC 8615, IETF, May 2019. Available:
https://www.rfc-editor.org/rfc/rfc8615

\subsubsection{Informative References}\label{informative-references}

{[}9{]} J. Cao and C. Arango Gutierrez, ``Agent Identity Protocol:
Agentic Authentication and Authorized Policy Enforcement,''
Internet-Draft draft-aip-agent-identity-protocol-00, IETF, March 2026.
Available:
https://datatracker.ietf.org/doc/html/draft-aip-agent-identity-protocol-00

{[}10{]} S. Ramachandran, N. Kothapalli, et al., ``Agentic JWT: A Secure
Delegation Protocol for Autonomous AI Agents,'' arXiv, 2025. Available:
https://arxiv.org/abs/2509.13603

{[}11{]} B. Zelenka et al., ``User-Controlled Authorization Network
(UCAN) Specification,'' Version 1.0.0-rc.1, 2024. Available:
https://github.com/ucan-wg/spec

{[}12{]} Biscuit Auth, ``Biscuit Authorization Token: Specification and
Documentation,'' accessed April 2026. Available:
https://www.biscuitsec.org/

{[}13{]} OASIS, ``eXtensible Access Control Markup Language (XACML)
Version 3.0,'' OASIS Standard, January 2013. Available:
https://www.oasis-open.org/committees/tc\_home.php?wg\_abbrev=xacml

{[}14{]} Open Policy Agent, ``Policy Language Reference,'' accessed
April 2026. Available:
https://www.openpolicyagent.org/docs/latest/policy-language/

{[}15{]} Cedar Policy, ``Cedar Policy Language Reference Guide,''
accessed April 2026. Available: https://docs.cedarpolicy.com/

{[}16{]} R. Pang, R. Caceres, M. Burrows, Z. Chen, P. Dave, N. Germer,
A. Golynski, K. Graney, N. Kang, L. Kissner, J. Korn, A. Parmar, C.
Richards, and M. Wang, ``Zanzibar: Google's Consistent, Global
Authorization System,'' in \emph{Proceedings of the 2019 USENIX Annual
Technical Conference (USENIX ATC '19)}, 2019. Available:
https://www.usenix.org/conference/atc19/presentation/pang

{[}17{]} Anthropic and contributors, ``Authorization - Model Context
Protocol,'' Model Context Protocol Specification, 2025. Available:
https://modelcontextprotocol.io/specification/2025-06-18/basic/authorization

{[}18{]} SPIFFE Project, ``SPIFFE ID Specification,'' accessed April
2026. Available: https://spiffe.io/docs/latest/spiffe-specs/spiffe-id/

{[}19{]} Google, ``Agent Discovery: The Agent Card,'' Agent2Agent (A2A)
Protocol Specification, accessed April 2026. Available:
https://google-a2a.github.io/A2A/latest/topics/agent-discovery/

{[}20{]} Project NANDA, ``Project NANDA Repository,'' accessed April
2026. Available: https://github.com/projnanda/projnanda

{[}21{]} OpenID Foundation, ``Authorization API 1.0,'' OpenID Final
Specification, January 2026. Available:
https://openid.net/specs/authorization-api-1\_0.html

{[}22{]} R. Iannella and S. Villata, ``ODRL Information Model 2.2,'' W3C
Recommendation, February 2018. Available:
https://www.w3.org/TR/odrl-model/

{[}23{]} W3C Credentials Community Group, ``Authorization Capabilities
for Linked Data v0.3,'' draft specification, accessed April 2026.
Available: https://w3c-ccg.github.io/zcap-spec/

{[}24{]} W3C Credentials Community Group, ``Incubation Specifications,''
accessed April 2026. Available:
https://w3c-ccg.org/specifications/incubation/

{[}25{]} OpenID Foundation, ``OpenID Shared Signals Framework
Specification 1.0,'' OpenID Final Specification, August 2025. Available:
https://openid.net/specs/openid-sharedsignals-framework-1\_0-final.html

{[}26{]} AGNTCY, ``Building infrastructure for the Internet of Agents,''
accessed April 2026. Available: https://agntcy.org/

{[}27{]} S. Cecchetti, ``Cedar Profile for OAuth 2.0 Rich Authorization
Requests,'' Internet-Draft draft-cecchetti-oauth-rar-cedar-02, IETF,
February 2024. Available:
https://www.ietf.org/archive/id/draft-cecchetti-oauth-rar-cedar-02.html

{[}28{]} T. South, S. Marro, T. Hardjono, R. Mahari, C. D. Whitney, D.
Greenwood, A. Chan, and A. Pentland, ``Authenticated Delegation and
Authorized AI Agents,'' arXiv:2501.09674, January 2025. Available:
https://arxiv.org/abs/2501.09674
\endgroup

\end{document}